\documentclass[twocolumn,showpacs,preprintnumbers,amssymb]{revtex4}

\usepackage{graphicx}% Include figure files
\usepackage{epsfig}
\usepackage{dcolumn}% Align table columns on decimal point
\usepackage{bm}% bold math
\def\gsim{\lower0.5ex\hbox{$\:\buildrel >\over\sim\:$}}
\def\lsim{\lower0.5ex\hbox{$\:\buildrel <\over\sim\:$}}

\def \n{\noindent}

\begin{document}

\preprint{AMES-HET-02-02}   
\preprint{BNL-HET-02/1}

\title{Flavor changing Z-decays from scalar interactions at a Giga-Z Linear Collider} 

\author{David Atwood}%
\email{atwood@iastate.edu}
\affiliation{Department of Physics and Astronomy, Iowa State University, Ames,
IA 50011, USA}
\author{Shaouly Bar-Shalom}%
\email{shaouly@physics.technion.ac.il}
\affiliation{Theoretical Physics Group, Rafael, Haifa 31021, Israel}
\author{Gad Eilam}
\email{eilam@physics.technion.ac.il}
\affiliation{Physics Department, Technion-Institute of Technology, Haifa 32000, Israel}
\author{Amarjit Soni}%
\email{soni@bnl.gov}
\affiliation{Theory Group, Brookhaven National Laboratory, Upton, NY 11973, USA}
\date{\today}

\begin{abstract}
The flavor changing decay $Z \to d_I \bar d_J$ is investigated with
special emphasis on the $b \bar s$ final state. 
Various models for flavor violation are considered: 
two Higgs doublet models (2HDM's), 
supersymmetry (SUSY) with flavor violation in the up 
and down-type squark mass matrices and SUSY with flavor violation 
mediated by $R$-parity-violating interaction. 
We find that, within the SUSY scenarios for flavor violation, 
the branching ratio for the decay $Z \to b \bar s$ can reach 
$10^{-6}$ for large $\tan\beta$ values, 
while the typical size for this branching ratio in the 2HDM's
considered is about two orders of magnitudes smaller at best.
Thus, flavor changing SUSY signatures in radiative 
$Z$ decays such as $Z\to b \bar s$ may be accessible to 
future ``Z factories'' such as a Giga-Z version of the TESLA design.
\end{abstract}


\maketitle

%
%
%
%%%%%%%%%%%%%%%%%%%%%%%%%%%%%%%%%%%%%%%>>>begin main text
%
%
%

\section{Introduction}
 
Rare processes involving various particles have always been a gold mine for
extracting interesting physics \cite{isidori,mele}. For example, the smallness
of Flavor Changing Neutral Currents (FCNC) in the $K$ system prompted
the introduction of the GIM mechanism and subsequently to the
prediction  $m_c\approx 1.5$ GeV and the discovery
of $J/\Psi$ and $D$'s. $B \bar{B}$ mixing was a precursor to a heavy top 
quark, as confirmed by experiment. FCNC 
rare top decays, for which there are only weak
upper bounds, will hopefully be discovered in future experiments,
thus serving as direct indications for deviations from the Standard 
Model (SM), since the latter leads to branching ratios which are 
smaller than $10^{-10}$. 

The situation in rare $Z$ decays, which is the subject of this 
paper, bears some similarities to rare $t$ decays. In both cases
the SM results from the loop induced FCNC decays are very small,
beyond reach, at least for $t$,
in the foreseeable future. Therefore, any significant 
detection of a rare decay at the level higher than $10^{-10}$ or $10^{-8}$
for $t$ or $Z$, respectively, would serve as an indisputable proof for physics
beyond the the SM. If new physics is ``around the corner'',
{\it i.e.} at $\approx 1$ TeV, the $Z$ boson and the top
being so close to that scale, are expected to be
the particles most affected by new physics.

In this paper we study the rare decays $Z \to d_I \bar{d}_J$,
where $I,~J$ indicate the generation index of a charge $-1/3$
quark, in various models. 
In the SM it was found that \cite{SM} ${\rm Br}(Z \to b\bar{s}+
\bar{b}s) \sim 10^{-8}$; we do not repeat the SM calculation here. 
Three of the models we discuss, have already been considered
in connection with $Z\to b\bar{s}$, namely the 2 Higgs Doublets
Model type II (2HDMII) \cite{2HDMIIref}, Supersymmetry (SUSY)
\cite{bsmixold} and SUSY with R-Parity Violation (RPV) \cite{shemtob}.
Therefore we comment , wherever it is relevant in the coming sections,
about differences and similarities with previous works.
Note that in addition one can find in the literature discussion of
FCNC hadronic $Z$ decays, in models not covered by us in
the present article \cite{others}.
 
Experimentally, the attention devoted to FCNC in $Z$ decays
at LEP and SLD has been close to nil. The best upper limit is \cite{delphi}

\begin{equation}
\sum_{q=d,s}\rm{Br}(Z\to b {\bar q})\leq 1.8 \times 10^{-3}~\rm{at}~90\%~CL\ . 
\end{equation}

\noindent This preliminary result is 
based on about $3.5\times 10^6$ hadronic decays, and we used \cite{pdg}
$\rm{Br}(Z\to \rm{hadrons})=0.7$. We urge our experimental colleagues
to sift through their LEP data to improve the current, rather loose, limit.

In the future, there will be at least two venues in which $Z$ bosons will
be produced in much larger quantities than their number in LEP.
In a high luminosity LHC with an integrated luminosity of $100~\rm{fb}^{-1}$, 
one expects $5.5\times 10^9$ Z bosons to be produced \cite{ZinLHC}. A cleaner
environment for the processes at hand, will be provided by a future
$e^+ e^-$ linear collider. In particular, there is a viable possibility
to lower the TESLA center of mass energy down to $\sqrt{s}=m_Z$, the
so called ``Giga-Z'' option. With integrated luminosity of $30~\rm{fb}^{-1}$,
it is possible to produce more than $10^9$ $Z$ bosons \cite{ZinTESLA},
about 2 orders of magnitudes larger than in LEP. To grasp the improvement
in going from LEP to Giga-Z option of TESLA, we note that while 
the sensitivity of LEP to $Z\to\tau\mu$ was $\approx 10^{-5}$
\cite{pdg}, it is expected to be $\approx 10^{-8}$ in Giga-Z TESLA
\cite{ZinTESLA}.

Beyond LHC and the $e^+ e^-$ linear collider, there is also considerable 
interest in the community for a high energy muon collider \cite{mucol1}.
If this ever becomes a reality, it would also afford another very good 
opportunity for studying rare flavor changing decays and interactions 
\cite{mucol2}
                            
The paper is organized as follows: In Section 2 we present a generic 
calculation of the $Z d_I {\bar d}_J$ vertex at the one loop level.
This result will assist us to evaluate the branching ratio for the FCNC
$Z$ decays in any particular model. In Section 3, the results of 
two variants of the Two Higgs Doublets Model (2HDM), 
namely the so called type II 2HDM (2HDMII) and 
the Top-Higgs Two Doublets Model (T2HDM),
will be reported.  In 2HDMII, T2HDM we get the disappointing 
results ${\rm Br}(Z \to b\bar{s}+\bar{b}s)\sim 10^{-10},~10^{-8}$, 
respectively.
We then move on to Section 4, where our results in Supersymmetry (SUSY)
with squark mixing, are displayed. Again, two options are presented,
the first one with ${\tilde b}-{\tilde s}$ mixing and the second
one with ${\tilde t}-{\tilde c}$ mixing. In the first case the 
branching ratio can reach a respectable 
${\rm Br}(Z \to b\bar{s}+\bar{b}s)\sim 10^{-6}$
while the second case yields a branching ratio of ${\cal O}(10^{-8})$.
In Section 5 we turn to SUSY with R-parity violation (RPV), 
where the effects
of $\lambda^{\prime}$ trilinear coupling terms in
the RPV superpotential and of $b$ terms ($b$ is the coefficient of the
soft breaking RPV bilinear term), 
are considered. Two categories of RPV are considered: Those which
lead to a branching ratio $\propto \lambda^{\prime}\times \lambda^{\prime}$
and those with a branching ratio $\propto b \lambda^{\prime}$.
For the first category we get typically branching ratios at the
level of $10^{-10}$, while for the second type of RPV,
we find an encouraging possibility of
${\rm Br}(Z \to b\bar{s}+\bar{b}s)\sim 10^{-6}$. Finally, in Section
6 we summarize our results.

\section{Generic scalar calculation}

In this section we outline the
generic framework for calculating 
the radiative one-loop flavor changing interaction vertex $V d_I \bar d_J$ 
with $I \ne J$ and $V=Z$ or $\gamma$. 
We define the one-loop amplitude for 
$V \to d_I \bar d_J$
in terms of 
form factors which are calculated for the complete set of one-loop 
diagrams that can potentially contribute to $V \to d_I \bar d_J$ 
in the presence of flavor violating interactions between neutral scalars and 
fermions as well as non-diagonal vertices of charged scalars with fermions of 
different generations.  

\begin{figure}
\includegraphics{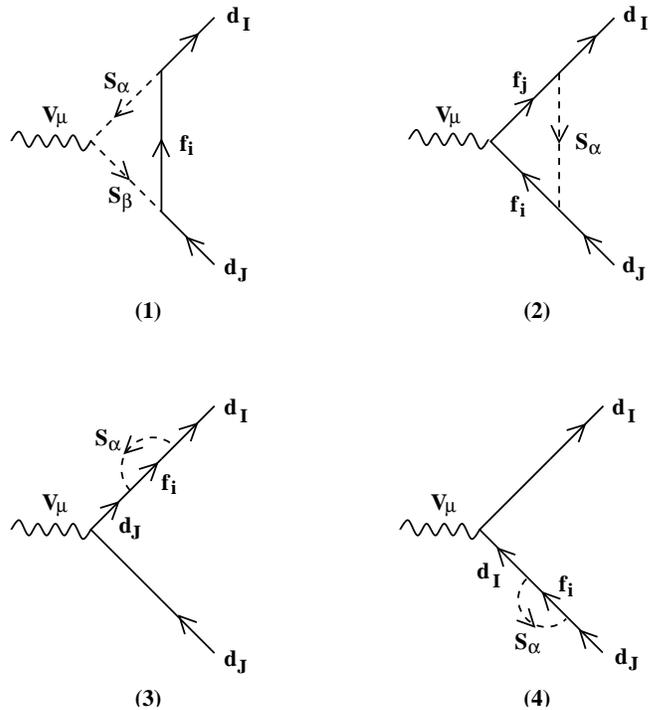}
\caption{\label{fig1} One-loop diagrams that contribute to the 
flavor changing transition $V \to d_I \bar d_J$, due to scalar-fermion 
exchanges.}
\end{figure}

The diagrams that modify (at one loop)
the $Vd_I \bar d_J$ coupling due to charged or neutral 
scalar exchanges are depicted
in Fig.~\ref{fig1}.

In what follows, we will denote the internal scalars ($S$) 
in the loops 
by Greek letters and fermions ($f$) will be given 
Latin indices $i,j$. 

%\begin{figure}[htb]
%\psfull
% \begin{center}
%  \leavevmode
%  \epsfig{file=fig1.eps,height=8cm,width=10cm,bbllx=0cm,bblly=2cm,bburx=20cm,bbury=25cm,angle=0}
% \end{center}
%\caption{\emph{One-loop diagrams that contribute to the 
%flavor changing transition $V \to d_I \bar d_J$, due to scalar-fermion 
%exchanges.}}
%\label{fig1}
%\end{figure}

From Fig.~\ref{fig1} it is evident that we have only three types 
of interaction vertices to consider. These are defined as follows:

%\newpage
\begin{enumerate}

\item \underline{$V_\mu - f_i - \bar f_j$ interaction}

\vspace{-3.0cm}\vbox to 2in{\epsfxsize=2 in \hspace{-3.0cm}\vspace{0.7cm}\epsfbox[-100 -500 100 300]{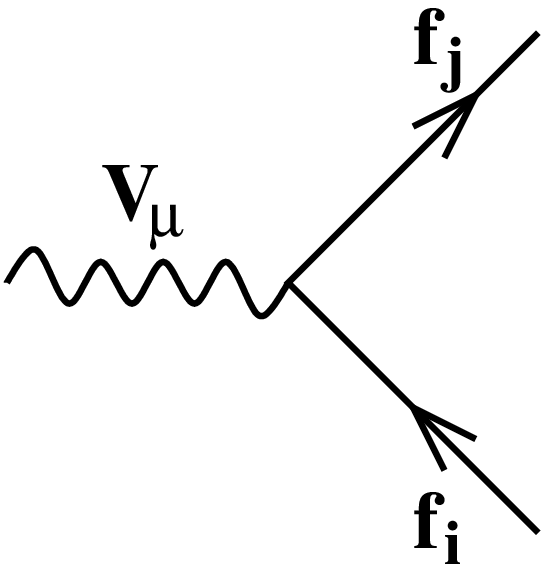}}
\hspace{3.0cm}\vspace{-0.79cm}$i \gamma_\mu \left( a_{L(Vf)}^{ij} L +
 a_{R(Vf)}^{ij} R \right) $
\begin{eqnarray}
~ \label{Vff}
\end{eqnarray}

\vspace{2.0cm}

\noindent where $L(R)=(1-(+)\gamma_5)/2$. For the case of
the SM couplings of a vector boson $V$ to a pair of fermions, i.e., 
$f=u$ (up-quark) or $f=d$ (down-quark), there are only 
diagonal $Vff$ couplings. In this case we have:

%\begin{eqnarray}
%a_{L(Vf)}^{ij} (i=j) =  a_{L(Vf)} ~~,~~ a_{R(Vf)}^{ij} (i=j) =  a_{R(Vf)} 
%\label{Vdd} ~.
%\end{eqnarray}

\begin{eqnarray}
a_{L,R(Vf)}^{ij} (i=j) \equiv  a_{L,R(Vf)} \label{Vdd} ~.
\end{eqnarray}

\noindent where

\begin{eqnarray}
&& a_{L,R(Zf)} = -g_Z \left(T_{L,R}^{3(f)}-s_W^2Q_f\right) \label{eq3}~, \\
&& a_{L(\gamma f)} = a_{R(\gamma f)} = -g_\gamma Q_f  \label{vffSM}~,
\end{eqnarray}

\noindent with $T^{3(u)}_L=1/2$ and $T^{3(d)}_L=-1/2$ for an up and 
down-quark, 
respectively, 
and $T^{3(f)}_R=0$. Also, $Q_f$ is the charge of $f$ and

\begin{eqnarray}
g_Z=\frac{e}{s_Wc_W} ~~,~~g_\gamma=e \label{Vddlast}~.
\end{eqnarray}

\item \underline{$V_\mu - S_\alpha - S_\beta$ interaction}

\vspace{-3.0cm}\vbox to 2in{\epsfxsize=2 in \hspace{-3.0cm}\vspace{0.7cm}\epsfbox[-100 -500 100 300]{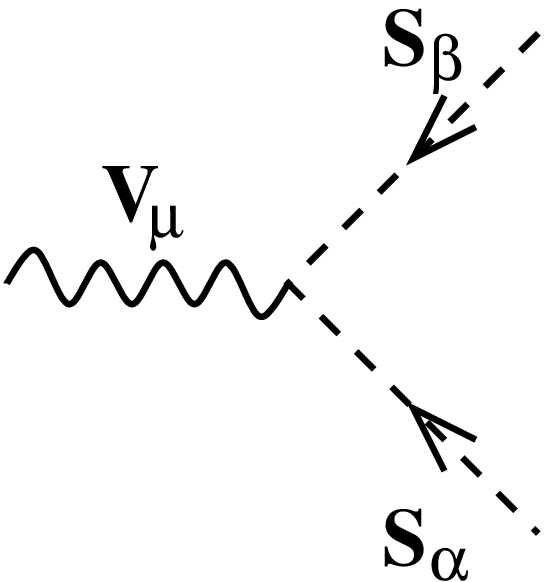}}
\hspace{3.0cm}\vspace{-0.79cm}$i g_V^{\alpha \beta} 
\left( p_\alpha - p_\beta \right)_\mu $
\begin{eqnarray}
~ \label{VSS}
\end{eqnarray}

\vspace{2.0cm}

\noindent where $S_\alpha$ and $S_\beta$ are 
charged or neutral spin 0 particles with incoming 4-momenta 
$p_\alpha$ and $p_\beta$, respectively.

\item \underline{$S_\alpha - \bar f_i - d_j $ interaction}

%\vspace{-3.0cm}\vbox to 2in{\epsfxsize=2 in \vspace{2.7cm}\epsfbox[-100 -100 100 200]{fr3.eps}}
%\hspace{6.0cm}\vspace{-1.0cm}$i \left(
%b_{L(\alpha)}^{ij} L + b_{R(\alpha)}^{ij} R \right) $
%\begin{eqnarray}
%~ \label{Sdf}
%\end{eqnarray}
%
%\vspace{3.0cm}

\vspace{-3.0cm}\vbox to 2in{\epsfxsize=2 in \hspace{-3.0cm}\vspace{0.7cm}\epsfbox[-100 -500 100 300]{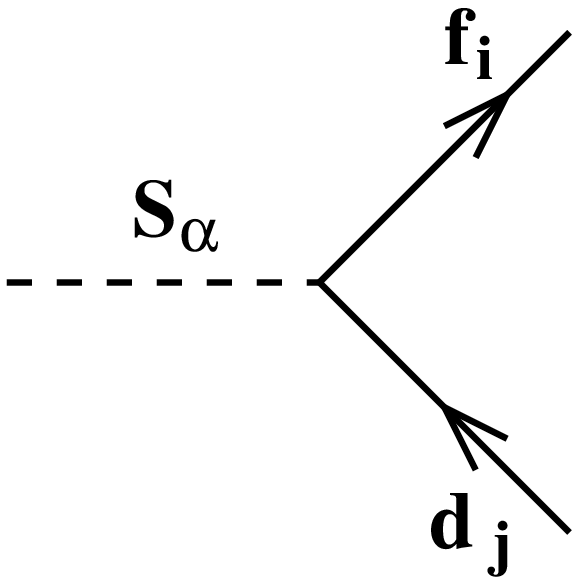}}
\hspace{3.0cm}\vspace{-0.89cm}$i \left(
b_{L(\alpha)}^{ij} L + b_{R(\alpha)}^{ij} R \right) $
\begin{eqnarray}
~ \label{Sdf}
\end{eqnarray}

\vspace{2.0cm}

\noindent where $d$ is a down-quark.

\end{enumerate}

The one-loop amplitudes ${\cal M}_k$ ($k=1,2,3,4$ corresponding
to diagrams $1,2,3,4$ in Fig.~\ref{fig1}, respectively), for the 
decay $V \to d_I \bar d_J$ can be parametrized generically as follows:

\vspace{2.5cm}

\begin{widetext}
\begin{eqnarray}
{\cal M}_k^{IJ} \equiv \frac{i}{16 \pi^2} 
\epsilon^V_\mu (q) \bar u_{d_I} (p_I)
\left\{ \gamma^\mu \left( A^{IJ}_{L,k} L + A^{IJ}_{R,k} R \right) +
\left( B^{IJ}_{L,k} L + B^{IJ}_{R,k} R \right) p_I^\mu \right\} v_{\bar d_J} (p_J) 
\label{muk}~,
\end{eqnarray}
\end{widetext}

\noindent where $\epsilon^V(q)$ is the polarization vector of $V$, $q$ is 
its 4-momenta and $\bar u_{d_I}~(v_{\bar d_J})$ is the Dirac spinor of the 
outgoing $d_I$ with 4-momenta $p_I$ ($\bar d_J$ with 4-momenta $p_J$) such 
that $q=p_I+p_J$.
Also, $A_{L,k}^{IJ},~A_{R,k}^{IJ},~B_{L,k}^{IJ},~B_{R,k}^{IJ}$ are 
momentum dependent form factors.

Using the Feynman rules in (\ref{Vff}), (\ref{VSS}) and (\ref{Sdf}), these 
form factors can be readily calculated for each diagram. 
Neglecting terms of ${\cal O}(m_b/\sqrt{q^2})$ we get:

\begin{eqnarray}
A^{IJ}_{L,1} &=& -2 \sum_{\alpha,\beta,i} 
g_V^{\alpha \beta} b_{L(\alpha)}^{iI} 
 b_{L(\beta)}^{iJ} C_{24}^{1} \label{firstamp} ~,  \\
B^{IJ}_{L,1} &=& 2 \sum_{\alpha,\beta,i} {\hat P}_i m_{f_i} 
g_V^{\alpha \beta} b_{R(\alpha)}^{iI} 
 b_{L(\beta)}^{iJ} \left( C_{0}^{1}+ 
C_{11}^{1} \right) ~, \\
A^{IJ}_{L,2} &=& - 2 \sum_{\alpha,i,j} b_{L(\alpha)}^{iI}  b_{L(\alpha)}^{jJ} 
\left\{ a_{R(Vf)}^{ij}  
\left[ p_I \cdot p_J  \left( C_{23}^{2} - C_{22}^{2} 
\right)  \right. \right. \nonumber \\
&& \left. \left. - C_{24}^{2} \right] + {\hat P}_i {\hat P}_j 
m_{f_i} m_{f_j} a_{L(Vf)}^{ij} 
C_0^{2} \right\} ~, \\
B^{IJ}_{L,2} &=& -2 \sum_{\alpha,i,j}  b_{R(\alpha)}^{iI}  b_{L(\alpha)}^{jJ} 
\left\{ 
{\hat P}_i m_{f_i} a_{L(Vf)}^{ij}  
 \left( C_{11}^{2} - C_{12}^{2} 
\right) \right. \nonumber \\
&& \left. + {\hat P}_j m_{f_j} a_{R(Vf)}^{ij}  
C_{12}^{2} \right\} ~, 
\end{eqnarray}

\noindent where ${\hat P}_i=-1$ if the internal fermion in the loop
is a charged conjugate
state ($f_i^c$) or else ${\hat P}_i=1$.

Combining the contribution of the two self energy diagrams, i.e., 
${\cal M}_3+{\cal M}_4 \equiv {\cal M}_{34}$, and similarly for the form 
factors, e.g., 
$A^{IJ}_{L,3} + A^{IJ}_{L,4} \equiv A^{IJ}_{L,34}$ etc., we have:

\begin{eqnarray}
%A^{IJ}_{L,34} &=& a_{L(Vd)} \sum_{\alpha,i} \left\{ 
% b_{L(\alpha)}^{iI}  b_{L(\alpha)}^{iJ} B_1^{3} + 
%\right. \nonumber \\ 
%&& \left. \frac{1}{m_I^2 - m_J^2}  
%\left[ m_J \left(B_1^{3} - B_1^{4} \right)
%\left( m_J b_{L(\alpha)}^{iI}  b_{L(\alpha)}^{iJ} 
%  +  m_I b_{R(\alpha)}^{iI}  b_{R(\alpha)}^{iJ} \right) +
% \right. \right. \nonumber \\
%&&  \left. \left. P_i m_{f_i} \left(B_0^{4} - B_0^{3} \right)
%\left( m_I b_{R(\alpha)}^{iI}  b_{L(\alpha)}^{iJ} +
%m_J b_{L(\alpha)}^{iI}  b_{R(\alpha)}^{iJ} \right) \right] \right\} 
%~,\\
A^{IJ}_{L,34} &=& a_{L(Vd)} \sum_{\alpha,i}
b_{L(\alpha)}^{iI}  b_{L(\alpha)}^{iJ} B_1^{3} ~,\\
B^{IJ}_{L,34} &=& 0 \label{lastamp}~.
\end{eqnarray}
 
\noindent The right-handed form factors, 
$A^{IJ}_{R,k}$ and $B^{IJ}_{R,k}$, are obtained from the corresponding 
left handed ones, 
$A^{IJ}_{L,k}$ and $B^{IJ}_{L,k}$ respectively,  
by interchanging $L \to R$ and $R \to L$ in all the 
couplings which appear in (\ref{firstamp})-(\ref{lastamp}).

The three-point one-loop form factors $C_x^k$ with 
$x \in 0,11,12,21,22,23,24$, and two-point form factors 
$B_x^k$ with $x \in 0,1$, 
correspond to the loop 
integrals of diagrams $k=1-4$ and are given by: 

\begin{eqnarray}
C_x^{1} &=& 
C_x \left(m_{f_i}^2,m_{S_\alpha}^2,m_{S_\beta}^2,m_{d_I}^2,q^2,m_{d_J}^2 
\right) ~,\\
C_x^{2} &=& 
C_x \left(m_{S_\alpha}^2,m_{f_i}^2,m_{f_j}^2,m_{d_J}^2,q^2,m_{d_I}^2 
\right) ~,\\
B_x^{3} &=& B_x \left(m_{f_i}^2,m_{S_\alpha}^2,m_{d_I}^2 \right) ~,
\end{eqnarray}

\noindent where $B_x(m_1^2,m_2^2,p^2)$ and 
$C_x(m_1^2,m_2^2,m_3^2,p_1^2,p_2^2,p_3^2)$ are defined in the appendix.

\noindent In terms of the above form factors, the partial 
width for the decay $Z \to d_I \bar d_J$ is:

\begin{widetext}
\begin{eqnarray}
\Gamma(Z\to d_I \bar d_J) = 
\frac{N_C}{3} \left( \frac{1}{16 \pi^2} \right)^2 
\frac{M_Z}{16 \pi} \times
\left[ 2 \left( \mid A_L^T \mid^2 + \mid A_R^T \mid^2 \right) 
+\frac{M_Z^2}{4}  \left( \mid B_L^T \mid^2 + \mid B_R^T \mid^2 \right) 
\right] \label{ampl}~,
\end{eqnarray}
\end{widetext}

\noindent where $N_C=3$ is the color factor and

\begin{eqnarray}
A_{P}^T &\equiv& A_{P,1}^{IJ} +  A_{P,2}^{IJ} + A_{P,34}^{IJ} \label{apt}~,\\
B_{P}^T &\equiv& B_{P,1}^{IJ} +  B_{P,2}^{IJ} + B_{P,34}^{IJ} \label{bpt}~,
\end{eqnarray}

\noindent for $P=L$ and $R$.

\section{Two Higgs doublets models}

In a 2HDM with flavor diagonal couplings of the neutral Higgs 
to down-quarks, the flavor changing decay $Z \to d_I \bar d_J$ proceeds 
through the one-loop 
diagrams in Fig.~\ref{fig1}.  

The interaction vertices required for the calculation of 
the form factors defined in (\ref{muk}) in such models are:
 
\begin{eqnarray}
V_\mu f_i \bar f_j &\to& Z_\mu u_i \bar u_j ~,\nonumber \\
V_\mu S_\alpha S_\beta &\to& Z_\mu H^+ H^- \label{2hdmint}~, \\
S_\alpha \bar f_i d_j &\to& H^+ \bar u_i d_j ~, \nonumber
\end{eqnarray}

\noindent where the 
$Z_\mu u_i \bar u_j$ coupling is the SM one as given in 
(\ref{Vdd})-(\ref{Vddlast}), $S_1=H^+$ is the only 
charged Higgs present in this type of models and 
$f_i=u_i$, $i=1,2,3$ for the three up-type quarks $u_1=u$, $u_2=c$, 
$u_3=t$. 

The $Z_\mu H^+ H^-$ coupling is obtained from the scalar kinetic term 
$\left(D^\mu \Phi_i \right)^\dagger \left(D_\mu \Phi_i \right)$,
where $\Phi_{1,2}$ are the two SU(2) Higgs doublets. 
This coupling is, therefore,  
generic to any version of a 2HDM.   

The coupling $H^+ \bar u_i d_j$ is obtained from the Yukawa potential.
The most general Yukawa interaction of a 2HDM can be written as 
(see e.g., \cite{yukawaref}):

\begin{eqnarray} 
{\cal L}_Y &=& - \sum_{i,j}  \bar Q_L^i 
\left[ \left(U_{ij}^1 \tilde\Phi_1  +U_{ij}^2 \tilde\Phi_2 \right) u^j_R 
\right. \nonumber\\
&& \left. + 
\left(D_{ij}^1 \Phi_1  +D_{ij}^2 \Phi_2 \right) d^j_R \right] \label{yukawa}~,
\end{eqnarray}

\noindent where $Q_L$ is the SU(2) left-handed quark doublet,
$u_R$ and $d_R$ are the right-handed up and down SU(2) singlets, 
respectively, and   
$\tilde\Phi_{1,2}=i \sigma_2 \Phi_{1,2}^*$. Also, $U^1,U^2,D^1,D^2$ 
are general Yukawa $3 \times 3$ matrices in flavor space.
The different types of 2HDM's are then categorized according to 
the choice of the Yukawa matrices $U^1,U^2,D^1,D^2$.

In what follows we will focus on two specific versions of a 2HDM:

\begin{itemize}

\item {\bf 2HDM of type II (2HDMII)}

The 2HDMII follows from the choice $U^1=0$ and $D^2=0$ 
in which case only $\Phi_2$ generates the masses of the up-type quarks 
and only $\Phi_1$ is responsible for the mass generation of 
the down-type quarks \cite{froggat}. This version of a 2HDM 
is realized in the Minimal Supersymmetric Standard Model (MSSM).

\item {\bf 2HDM ``for the top-quark'' (T2HDM)}

In the T2HDM \cite{das}, the large mass of the top-quark is 
accommodated in 
a natural fashion by coupling the second Higgs doublet 
($\Phi_2$), which has a much larger   
vacuum expectation value (VEV), only to the top-quark;
all other quarks are coupled to the 1st Higgs doublet ($\Phi_1$).
This scenario is, therefore, realized by setting in (\ref{yukawa}):

\begin{eqnarray}
U^1_{ij} &\to& G_{ij} \times (\delta_{j1}+\delta_{j2}) ~,\nonumber \\
U^2_{ij} &\to& G_{ij} \times \delta_{j3} \label{u2ij} ~,\\
D^2_{ij} &\to& 0 \nonumber ~,
\end{eqnarray}

\noindent where $G$ is again an unknown Yukawa $3 \times 3$ matrix 
in quark flavor space.  

\end{itemize}

Using the Lagrangian pieces given above, 
we list in Table \ref{tab1} all the couplings required for the calculation 
of $\Gamma(Z \to d_I \bar d_J)$ in a 2HDMII and in a T2HDM.

\begin{table*}[htb]
\begin{center}
%\begin{tabular}{||c|c|c||}
\begin{tabular}{c|c|c}
%\hline
%%%%%%%%%%%%%%%%%%%%%%%%%%%%%%%%%%%%%%%%%%%%%%%%%%%%
 & 2HDMII & T2HDM \\ 

~ & ~ & \\
\hline
~ & ~ & \\
scalar ($S_{\alpha=1}$) & $H^+$ & $H^+$ \\
~ & ~ & \\
%\hline
fermion ($f_i$) & $u_i$, $i=1,2,3$ &  $u_i$, $i=1,2,3$ \\
%\hline
~ & ~ & \\
$a_{L(Zf)}^{ij}$ & $a_{L(Zu)}$ & $a_{L(Zu)}$ \\
%\hline
~ & ~ & \\
$a_{R(Zf)}^{ij}$ & $a_{R(Zu)}$  & $a_{R(Zu)}$ \\
%\hline
~ & ~ & \\
$b_{L(\alpha=1)}^{ij}$ & 
$Y_{ij} \times \frac{1}{\tan^2\beta} \frac{m_{u_i}}{m_t}$ & 
$Y_{ij} \times 
\left[\frac{\Sigma_{i k}^\dagger V^{kj}_{CKM}}{m_t V_{ij}} 
\left(1+\frac{1}{\tan^2\beta} \right) - 
\frac{m_{u_i}}{m_t} \right]$ \\
~ & ~ & \\
%\hline
$b_{R(\alpha=1)}^{ij}$ & $Y_{ij} \times \frac{m_{d_j}}{m_t}$ & 
$Y_{ij} \times  \frac{m_{d_j}}{m_t}$ \\
~ & ~ & \\
%\hline
$g_Z^{\alpha=1 \beta=1}$ & $- e \frac{1-2s_W^2}{2 s_Wc_W}$ & 
$- e \frac{1-2s_W^2}{2 s_Wc_W}$ \\
~ & ~ & \\
%\hline
\end{tabular} 
\caption{The couplings required for the calculation 
of $\Gamma(Z\to d_I \bar d_J)$ in a 2HDMII and a T2HDM. 
The couplings follow the notation used in the 
Feynman rules of (\ref{Vff}), (\ref{VSS}) and 
(\ref{Sdf}). Also, $a_{R(Zu)},~a_{L(Zu)}$ are given in 
(\ref{Vdd})-(\ref{Vddlast}).}  
\label{tab1}
\end{center}
\end{table*}

\noindent In Table \ref{tab1}, $s_W,~c_W$ are the sine and cosine 
of the weak-mixing angle $\theta_W$, $m_{u_i}=m_u,~m_c~,m_t$ for 
$i=1,2,3$, respectively, and  

\begin{eqnarray} 
Y_{ij} = -\frac{e}{\sqrt{2}s_W} \frac{m_t}{M_W} \tan\beta V^{ij}_{CKM} 
\label{yij} ~.
\end{eqnarray}

\n with $V_{CKM}$ the Cabibbo-Kobayashi-Maskawa (CKM) matrix 
and $\tan\beta \equiv t_\beta = v_2/v_1$ (we will loosely refer  
to the ratio $v_2/v_1$ either as $\tan\beta$ or $t_\beta$), 
where $v_2(v_1)$ is the VEV of $\Phi_2(\Phi_1)$.  
Also, the matrix $\Sigma$ introduced in Table \ref{tab1} 
is composed out of the unitary matrix that diagonalizes the right-handed
up-type quarks and the Yukawa matrix $U_{ij}^2$ defined in (\ref{u2ij}). It, 
therefore, arises from the specific structure 
of the Yukawa interactions in the T2HDM and   
can be parametrized as 
follows \cite{9810552} 
(neglecting the mass of the first generation up-quark):

\begin{eqnarray}
\Sigma = \pmatrix{
0 & 0 & 0  \cr
0 & m_c \varepsilon_{ct}^2 |\xi|^2 & m_c \varepsilon_{ct} \xi^\star \cr
0 & m_c \xi \sqrt{1-|\varepsilon_{ct} \xi|^2}  &  m_t 
\left( 1-|\varepsilon_{ct} \xi|^2 \right) } \label{sigma}~,
\end{eqnarray}

\noindent where $ \varepsilon_{ct} = m_c/m_t$ and $\xi$ is an unknown 
parameter (assumed here to be real) 
whose ``natural'' size is of ${\cal O}(1)$. 

Notice that the specific structure of the 
T2HDM's Yukawa potential does not give rise to 
tree-level flavor changing couplings between a neutral Higgs and a pair of 
down-quarks (while allowing for tree-level neutral Higgs-top-charm couplings).
Therefore, the decay $Z \to d_I \bar d_J$ is not affected at one-loop
by flavor changing neutral Higgs-quark interactions.  

\subsection{$Z \to b \bar s$ in 2HDMII}

The charged Higgs one-loop contribution to the decay $Z \to b \bar s$ 
in a 2HDMII was examined 
before in \cite{2HDMIIref}.
Here we wish to recapitulate the salient features of 
this decay. 

On the left side of 
Fig.~\ref{fig2hdm} we plot $BR(Z \to b \bar s + \bar b s)$ as 
a function of $\tan\beta$ for charged Higgs masses of 
$100$, $400$ and 600 GeV. 
As can be seen, $BR(Z \to b \bar s + \bar b s)$ is maximal 
for low $\tan\beta \sim {\cal O}(1)$ for which 
it is controlled by the top-quark Yukawa coupling which is 
$\propto 1/\tan\beta$. Thus, in this range the 
required flavor transition is mediated by $t \to s$ and the 
$BR(Z \to b \bar s + \bar b s)$ is, therefore, essentially 
proportional to $(m_t/\tan\beta)^4 \times (V^{tb}_{CKM})^2 (V^{ts}_{CKM})^2$
which is the square of the product of the $tbH^+$ and $tsH^+$ Yukawa 
couplings. 

At around $\tan\beta \sim 13$ there is a ``turning point'' at which 
the $BR(Z \to b \bar s + \bar b s)$ starts to increase with $\tan\beta$. 
At this point the contributions from the top and charm-quark loop exchanges 
become comparable, since the charm-quark effect being  
$\propto m_s^2 m_b^2 \tan^4\beta 
(V^{cs}_{CKM})^2 (V^{cb}_{CKM})^2$ (for 
$\tan\beta \gsim 13$ the charm-quark exchange
is dominated by the Yukawa couplings $b_R^{22}$ and $b_R^{23}$, see
Table \ref{tab1})
equals that of the top-quark.
As $\tan\beta$ is further increased ($\tan\beta > 13$) 
both the top and charm-quark loop
exchanges are dominated by the right-handed 
down-quark Yukawa couplings $b_R^{ij}$ in (\ref{Sdf}) (which is 
$\propto \tan\beta$) and are, 
therefore, comparable and increasing with $\tan\beta$. 

Note that the curves in Fig.~\ref{fig2hdm} for the 2HDMII scenario 
(the left side) pass through unrealistic 
values in the $\tan\beta - m_{H^+}$ plane. 
In particular, the decay $b \to s \gamma$ imposes strong constraints 
on the $\tan\beta - m_{H^+}$ plane \cite{2HDMIIlim}:  
$m_{H^+} \gsim 400$ GeV independent of $\tan\beta$.
Thus, if $\tan\beta=1$, then the largest allowed 
value for the $BR(Z \to b \bar s + \bar b s)$  
is $\sim 10^{-10}$ if $m_{H^+}$ lies close to its 
lower bound from $b \to s \gamma$. 
For even smaller values of $\tan\beta$, say $\tan\beta \lsim 0.5$,  
the $b \to s \gamma$ constraint requires $m_{H^+} \gsim 500$ GeV 
for which the $BR(Z \to b \bar s + \bar b s)$ is again 
$\lsim 10^{-10}$ in the 2HDMII.   

\subsection{$Z \to b \bar s$ in T2HDM}

The main difference between the T2HDM and the 2HDMII charged Higgs 
sectors lies in
the structure of the $c d_i H^+$ Yukawa interactions ($d_i=d,~s,~b$
for $i=1,~2,~3$ respectively).
In particular, while in both models the top Yukawa coupling 
to down-quarks, i.e., the $\bar t_R d_L H^+$ coupling, 
is $\propto m_t V^{td}_{CKM}/\tan\beta$, the charm-quark Yukawa coupling
is completely 
altered by the presence of the matrix $\Sigma$ in (\ref{sigma}). 
Specifically, the $\bar c_R b_L H^+$ coupling
is $\propto m_c \xi^\star V^{tb}_{CKM} \tan\beta$ and the 
$\bar c_R s_L H^+$ coupling is $\propto m_c V^{cs}_{CKM} \tan\beta$. These 
couplings are, therefore, a factor of $\tan^2\beta \times 
(V^{tb}_{CKM}/V^{cb}_{CKM})$ 
and $\tan^2\beta$, respectively, larger than in the 2HDMII if
$\xi$ is of ${\cal O}(1)$.

\begin{figure*}[htb]
\psfull
 \begin{center}
  \leavevmode
  \epsfig{file=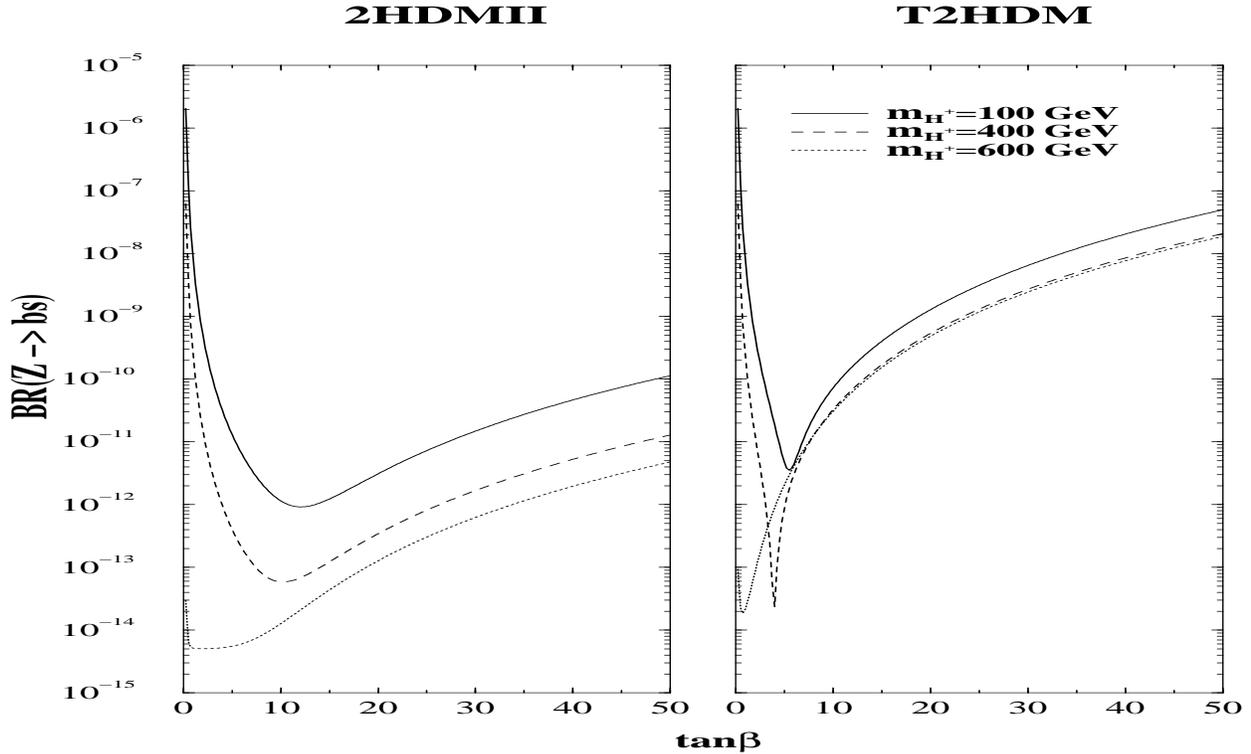,height=9cm,width=17cm,bbllx=0cm,bblly=2cm,bburx=20cm,bbury=25cm,angle=0}
 \end{center}
\caption{$BR(Z\to b \bar s + \bar b s)$ as a function 
of $\tan\beta$, for $m_{H^+}=100,~400$ and $600$ GeV, in 
a 2HDMII (left side) and a T2HDM with $\xi=1$ (right side).}
\label{fig2hdm}
\end{figure*}

As can be seen 
from Fig.~\ref{fig2hdm} (the right side), 
in the range $\tan\beta \lsim 5$ the $BR(Z \to b \bar s + \bar b s)$ 
is practically identical in both the T2HDM and the 2HDMII; 
in this range it is dominated by the top-quark loop 
and, therefore, by the top-quark Yukawa couplings to the 
$b$ and $s$-quarks which are 
essentially the same in these two versions of a 2HDM.
On the other hand, for larger values of $\tan\beta$, 
in contrast to the case of a 
2HDMII, in the T2HDM the charm-quark loop starts to dominate
due to the enhancement in the $\bar c_R b_L H^+$ and 
$\bar c_R s_L H^+$ Yukawa couplings 
(see discussion above).
In fact, the $\bar c_R b_L H^+$ coupling is doubly 
enhanced in the T2HDM; first by the $\tan\beta$ factor and second 
by a factor of $V^{tb}_{CKM}/V^{cb}_{CKM}$, i.e., in this model this coupling 
does not suffer from the CKM suppression factor $V^{cb}_{CKM}$.  

It should be emphasized that a large $\tan\beta$, e.g., 
$\tan\beta \gsim {\cal O}(10)$, is the ``working assumption'' 
of the T2HDM. In particular, the T2HDM looses its attractiveness 
in the small $\tan\beta$ regime, since in this
range it fails to explain the large top-quark mass - this being 
the main motivation
behind this version of a 2HDM.  
At the same time, taking into account low-energy experimental data from 
$K - \bar K$ mixing, $\epsilon_K$ and 
$b \to s \gamma$, the $\tan\beta - m_{H^+}$ plane 
is also constrained in the T2HDM \cite{9810552,t2hdmlim}, 
especially in the large $\tan\beta$
region in which this model differs from the usual 2HDMII.
For example, for $\xi=1$ and taking the SM best fit value for the 
Wolfenstein parameters $\rho$ and $\eta$, then the $\epsilon_K$ constraint
implies $m_{H^+} \gsim 500$ GeV for $\tan\beta \sim 20$ 
and $m_{H^+} \gsim 4$ TeV for $\tan\beta \sim 50$
\cite{t2hdmlim}.  
Imposing these bounds we find that 
$BR(Z \to b \bar s + \bar b s) \sim {\cal O}(10^{-8})$ is the best case
value in this model assuming a large $\tan\beta$.

\section{Supersymmetry with squark mixings}

In SUSY flavor changing phenomena  
can emanate from mixings of squarks of different generations through 
the soft breaking Lagrangian terms in the squark sector:

\begin{widetext}
\begin{eqnarray}
{\cal L}_{soft}^{squark}=
-\tilde Q_i^\dagger (M_Q^2)_{ij} \tilde Q_j
-\tilde U_i^\dagger (M_U^2)_{ij} \tilde U_j
-\tilde D_i^\dagger (M_D^2)_{ij} \tilde D_j
+ A_u^{ij} \tilde Q_i H_u \tilde U_j +A_d^{ij} \tilde Q_i H_d \tilde D_j 
\label{soft}~,
\end{eqnarray}
\end{widetext}

\noindent where $\tilde Q$ is the SU(2) scalar doublet and 
$\tilde U,~\tilde D$
are the  up-squark, down-squark SU(2) singlets, respectively. Also, 
$i,j$ are generation indices.
  
The mass matrices in the up-squark and down-squark sectors 
may then be generically expressed as:

\begin{eqnarray}
M_{\tilde U,\tilde D}^2 = 
\pmatrix{
(m_{\tilde U,\tilde D}^2)_{LL} & (m_{\tilde U,\tilde D}^2)_{LR} \cr
(m_{\tilde U,\tilde D}^2)_{LR}^\dagger & (m_{\tilde U,\tilde D}^2)_{RR} }
\end{eqnarray}

\noindent where 
$(m_{\tilde U,\tilde D}^2)_{LL}$, $(m_{\tilde U,\tilde D}^2)_{RR}$
and $(m_{\tilde U,\tilde D}^2)_{LR}$ are $3\times 3$ matrices 
in squark flavor space. 
In the super CKM basis the squark fields are 
rotated ``parallel'' to their fermionic super-partners. 
In this basis, and assuming that there is 
a typical common mass scale for the squarks, $m_0$, which is
sufficiently heavier than the electroweak mass scale 
($m_0^2/M_Z^2 \gg 1$), these matrices are related to 
the soft breaking 
bilinear and trilinear terms in (\ref{soft}) via \cite{misiak1}: 

\begin{widetext}
\begin{eqnarray}
(m_{\tilde U}^2)_{LL} \equiv V_L^U M_Q^2 V_L^{U \dagger}~&,&~
(m_{\tilde D}^2)_{LL} \equiv V_L^D M_Q^2 V_L^{D \dagger} ~,\nonumber \\
(m_{\tilde U}^2)_{RR} \equiv V_R^U M_U^{2T} V_R^{U \dagger}~&,&~
(m_{\tilde D}^2)_{RR} \equiv V_R^D M_D^{2T} V_R^{D \dagger}~,\\
(m_{\tilde U}^2)_{LR} \equiv - \frac{v \sin\beta}{\sqrt{2}} 
V_L^U A_u^* V_R^{U \dagger} ~&,&~
(m_{\tilde D}^2)_{LR} \equiv \frac{v \cos\beta}{\sqrt{2}} 
V_L^D A_d^* V_R^{D \dagger}~,\nonumber
\end{eqnarray}
\end{widetext}

\noindent where $V_{L,R}^D$ and $V_{L,R}^U$ are the rotation matrices that 
diagonalize the down and up-quark fermion mass matrices, respectively 
(the CMK matrix is $V_{CKM} = V_L^U V_L^{D \dagger}$). 

Assuming that flavor changing squark mixings are significant 
only in transitions 
between the third and second generation squarks, 
we choose the following textures for the $3\times 3$ matrices 
$(m_{\tilde U,\tilde D}^2)_{LL}$ and 
$(m_{\tilde U,\tilde D}^2)_{RR}$:   

\begin{eqnarray}
(m_{\tilde U,\tilde D}^2)_{LL,RR}=
\pmatrix{
1 &   0   &   0 \cr
0 &   1   &   \delta^{U,D(23)}_{LL,RR}  \cr
0 &  \delta^{U,D(32)}_{LL,RR} &  1 } m_0^2 \label{deltadef}~,
\end{eqnarray}

\noindent where $\delta^{U(23)}_{LL}$, $\delta^{U(32)}_{LL}$
$\delta^{U(23)}_{RR}$, $\delta^{U(32)}_{RR}$
and $\delta^{D(23)}_{LL}$, $\delta^{D(32)}_{LL}$
$\delta^{D(23)}_{RR}$, $\delta^{D(32)}_{RR}$ represent 
flavor mixings in the $\tilde t - \tilde c$ and
$\tilde b - \tilde s$ sectors, respectively. These 
flavor violating quantities 
emanate from non-diagonal entries   
in the bilinear soft breaking terms $M_Q^2,~M_U^2$
and $M_D^2$ in (\ref{soft}).
Similarly, $\delta^{U(23)}_{LR}$, $\delta^{U(32)}_{LR}$ and 
$\delta^{D(23)}_{LR}$, $\delta^{D(32)}_{LR}$ are associated with non-diagonal 
(flavor changing) entries in the trilinear soft breaking terms $A_u$ 
and $A_d$ defined in (\ref{soft}). As in 
\cite{yuan} we 
adopt the following simplified Ansatz:

\begin{eqnarray}
V_L^U A_u^* V_R^{U \dagger} = 
\pmatrix{
0 & 0 & 0 \cr
0 & 0 & x_u \cr
0 & y_u & 1 } A~, 
\end{eqnarray}

\n and
 
\begin{eqnarray}
V_L^D A_d^* V_R^{D \dagger}=
\pmatrix{
0 & 0 & 0 \cr
0 & 0 & x_d \cr
0 & y_d & 1 } A~, 
\end{eqnarray}  

\noindent such that $A$ is a common trilinear soft breaking parameter
for both up and down-squarks and the parameters $x_u,y_u$ and 
$x_d,y_d$ represent flavor mixing effects in 
the $\tilde t - \tilde c$ and
$\tilde b - \tilde s$ sectors, respectively.
It then follows that $\delta^{U(23)}_{LR}$, $\delta^{U(32)}_{LR}$ and 
$\delta^{D(23)}_{LR}$, $\delta^{D(32)}_{LR}$
are related to $x_u,~y_u$ and $x_d,~y_d$ via:

\begin{widetext}
\begin{eqnarray}
\delta^{U(23)}_{LR} = 
- x_{u} \frac{\sin\beta}{\sqrt{2}} \times \frac{v A}{m_0^2} ~~,~~
\delta^{U(32)}_{LR} = 
- y_{u} \frac{\sin\beta}{\sqrt{2}} \times  \frac{v A}{m_0^2} ~,\\
\delta^{D(23)}_{LR} = 
x_{d} \frac{\cos\beta}{\sqrt{2}} \times \frac{v A}{m_0^2} ~~,~~
\delta^{D(32)}_{LR} = 
y_{d} \frac{\cos\beta}{\sqrt{2}} \times  \frac{v A}{m_0^2} ~.
\end{eqnarray}
\end{widetext}

\n Within this mixing scenario, 
in which flavor changing effects in the squark
sector are present only in the second and third generations, 
the $6\times 6$ mass matrices in the up and down-squark sectors 
reduce to $4 \times 4$ matrices. 

For the $\tilde t - \tilde c$ sector, in the basis 
$\Phi^0_U = (\tilde c_L,~\tilde c_R,~\tilde t_L,~\tilde t_R)$, 
we then have:

\begin{equation}
{\tilde M}_{ct}^2 = 
\pmatrix{
1             & 0   & \delta^{U(23)}_{LL} & \delta^{U(23)}_{LR} \cr
0             & 1   & \delta^{U(32)}_{LR}  & \delta^{U(23)}_{RR} \cr
\delta^{U(32)}_{LL} & \delta^{U(32)}_{LR}  &  1  & -X_t/m_0^2 \cr
\delta^{U(23)}_{LR} & \delta^{U(32)}_{RR}  & -X_t/m_0^2 & 1 } m_0^2~, 
\end{equation}   

\noindent and similarly for the $\tilde b - \tilde s$ sector, in the basis 
$\Phi^0_U = (\tilde s_L,~\tilde s_R,~\tilde b_L,~\tilde b_R)$, 
we have:

\begin{equation}
{\tilde M}_{sb}^2 = \pmatrix{
1             & 0   & \delta^{D(23)}_{LL} & \delta^{D(23)}_{LR} \cr
0             & 1   & \delta^{D(32)}_{LR}  & \delta^{D(23)}_{RR} \cr
\delta^{D(32)}_{LL} & \delta^{D(32)}_{LR}  &  1  & -X_b/m_0^2 \cr
\delta^{D(23)}_{LR} & \delta^{D(32)}_{RR}  & -X_b/m_0^2 & 1 } m_0^2~.
\end{equation}   
  
\noindent The factors $X_t$ and $X_b$ are 
responsible for mixings between left and right handed
squarks of the same generation and are given by:

\begin{eqnarray}
X_t &=& \frac{v \sin\beta}{\sqrt{2}} A + \frac{m_t \mu}{\tan\beta} 
\label{xt}~,\\
X_b &=& -\frac{v \cos\beta}{\sqrt{2}} A + m_b \tan\beta \mu \label{xb} ~,
\end{eqnarray}

\noindent where $\mu$ is the usual Higgs mass parameter in the 
SUSY superpotential. 

After diagonalization of $\tilde M_{ct}^2$ and 
$\tilde M_{sb}^2$
one obtains the new mass-eigenstates 
which are now $\tilde t - \tilde c$ and $\tilde b - \tilde s$ ad-mixtures,
 respectively. The diagonalizing 
matrices $R_U$ and $R_D$ are defined through:

\begin{eqnarray}
\Phi^0_{U,i} = R_{U,ik} \Phi_{U,k}~~,~~ 
\Phi^0_{D,i} = R_{D,ik} \Phi_{D,k} \label{rotate}~,
\end{eqnarray}

\noindent where 

\begin{widetext}
\begin{eqnarray}
\Phi^0_U \equiv \pmatrix{ 
\tilde c_L \cr \tilde c_R \cr \tilde t_L \cr \tilde t_R } ~~,~~
\Phi_U \equiv \pmatrix{ 
\tilde c_1 \cr \tilde c_2 \cr \tilde t_1 \cr \tilde t_2 } ~~,~~
\Phi^0_D \equiv \pmatrix{ 
\tilde s_L \cr \tilde s_R \cr \tilde b_L \cr \tilde b_R } ~~,~~
\Phi_D \equiv \pmatrix{ 
\tilde s_1 \cr \tilde s_2 \cr \tilde b_1 \cr \tilde b_2 } ~,
\end{eqnarray}
\end{widetext}

\noindent and $\tilde u_{L,R},~\tilde d_{L,R}$ 
($u=c,t$ and $d=s,b$) are the SU(2) weak states, 
while 
$\tilde u_{1,2},~\tilde d_{1,2}$  
are the corresponding physical states (mass-eigenstates). 

Let us now consider separately the cases in which the one-loop flavor changing
decay $Z\to b \bar s$ is driven either by 
the $\tilde t - \tilde c$
or by the $\tilde b - \tilde s$ mixing phenomena. 

\subsection{\bf $\tilde b - \tilde s$ mixing}

Here the 
flavor changing decay $Z\to b \bar s$ is generated by 
one-loop exchanges of the $\tilde b - \tilde s$ 
admixture states, $\Phi_D$, and 
gluinos, $\tilde g$. 
We thus have $S_\alpha = \Phi_{D,\alpha}$ 
with $\alpha=1-4$,
and $f = \tilde g$ in the diagrams of Fig.~\ref{fig1}. 
Note that diagram (2) in Fig.~\ref{fig1}, which requires 
the $Vff$ coupling, does not contribute since 
the $Z$-boson does not couple to gluinos 
at tree-level.

The one-loop $\tilde b - \tilde s$ mixing effect 
on the decay $Z\to b \bar s$ was considered 
more than a decade ago in \cite{bsmixold}, 
where it was
assumed that flavor violation in the down-squark 
sector is controlled by 
radiative corrections to the down-squark mass matrix induced by off-diagonal
CKM elements.
Instead, as described above, 
we assume here that the flavor violation is rooted with 
non-diagonal entries in the soft SUSY breaking sector.
The approach taken here is, therefore, fundamentally different from 
the one suggested in \cite{bsmixold}.$^{[1]}$\footnotetext[1]{Note 
also that \cite{bsmixold} used unrealistic top-quark and squark masses.} 
%\footnotemark[1] 

The following interaction vertices are required for the calculation 
of $\Gamma(Z \to b \bar s)$ in the $\tilde b -\tilde s$ mixing scenario:

\begin{eqnarray}
V_\mu S_\alpha S_\beta &\to& Z_\mu \Phi_{D,\alpha}^\star \Phi_{D,\beta} ~, \\
S_\alpha \bar f_i d_j &\to& \Phi_{D,\alpha}^\star \bar{\tilde g} d_j ~.\nonumber
\end{eqnarray}

\noindent These are derived 
from \cite{rosiek}:

\begin{eqnarray}
{\cal L}(V_\mu \tilde d \tilde d) &=& -i
\left[ - \frac{1}{2} \frac{e}{s_W}
 A_\mu^3 + \frac{1}{6} \frac{e}{c_W} B_\mu \right] 
 \tilde d_{L,\ell}^{\star} \stackrel{\leftrightarrow}{\partial^\mu_{\tilde d}} 
\tilde d_{L,\ell} \nonumber \\
&&- i \frac{1}{3} \frac{e}{c_W}  B_\mu \tilde d_{R,\ell}
\stackrel{\leftrightarrow}{\partial^\mu_{\tilde d}} \tilde d_{R,\ell}^{\star}
\label{vdldl}~,
\end{eqnarray}
\begin{equation}
{\cal L}(\tilde d \tilde g d) = 
g_s \sqrt{2} T^a \bar{\tilde g} 
\left( -\tilde d_{L,\ell}^{\star} L +  
\tilde d_{R,\ell}^{\star} R \right) d_j +h.c.
~,
\end{equation}

\noindent where $g_s$ is the SU(3) coupling constant and $T^a$ 
are the SU(3) group generators.

Using the above interaction Lagrangian terms, the couplings 
required for the calculation of $\Gamma(Z \to b \bar s)$ in the form 
defined in (\ref{VSS}) and (\ref{Sdf}) are obtained by 
rotating the weak states,  
$\Phi^0_{D}$, to the physical 
states, $\Phi_{D}$, according to (\ref{rotate}). These couplings  
are given in Table \ref{tab2}.

\begin{table}[htb]
\begin{center}
%\begin{tabular}{||c|c||}
\begin{tabular}{c|c}
%\hline
%%%%%%%%%%%%%%%%%%%%%%%%%%%%%%%%%%%%%%%%%%%%%%%%%%%%
 & SUSY with $\tilde b - \tilde s$ mixing\\ 
~ \\
\hline
~ \\
scalar ($S_{\alpha}$) & $\Phi_{D,\alpha},~~\alpha=1,2,3,4$ \\
~ \\
%\hline
fermion ($f_i$) & $\tilde g$ \\
~ \\
%\hline
$a_{L(Zf)}^{ij}$ & 0 \\
~ \\
%\hline
$a_{R(Zf)}^{ij}$ & 0 \\
~ \\
%\hline
$b_{L(\alpha)}^{ij}$ & 
$- \sqrt{2} g_s T^a 
\left( R^{\star}_{D,3 \alpha} \delta_{3j} + 
R^{\star}_{D,1 \alpha} \delta_{2j} \right)$ \\
~ \\
%\hline
$b_{R(\alpha)}^{ij}$ & $- \sqrt{2} g_s T^a 
\left( R^{\star}_{D,4 \alpha} \delta_{3j} + 
R^{\star}_{D,2 \alpha} \delta_{2j} \right)$\\
~ \\
%\hline
$g_Z^{\alpha \beta}$ & $\frac{-e}{2s_W c_W} 
\left( R_{D,1 \alpha}^{\star} 
R_{D,1 \beta}
+  R_{D,3 \alpha}^{\star} R_{D,3 \beta} - 
\frac{2}{3} s_W^2 \delta_{\alpha \beta} 
\right) $ \\
~ \\
%\hline
\end{tabular} 
\caption{The couplings required for the calculation 
of $\Gamma(Z\to b \bar s)$ in the MSSM with $\tilde b - \tilde s$ mixing.
These couplings follow the notation used in the 
Feynman rules defined by (\ref{Vff}), (\ref{VSS}) and 
(\ref{Sdf}).  
\label{tab2}}
\end{center}
\end{table}

The relevant low-energy SUSY parameter space for the 
$\tilde b - \tilde s$ mixing scenario is characterized 
as follows:$^{[2]}$\footnotetext[2]{The term 
``low-energy'' refers to the electroweak (or collider energies) scale 
and 
is used in order to distinguish it from 
the scale 
in which the soft breaking couplings are generated (e.g., the GUT scale).}

\begin{description}

\item{I.} Some of the flavor changing parameters in the 
$\tilde b - \tilde s$ sector
are severely constrained by the $b \to s \gamma$ decay 
\cite{misiak1,deltalimits}.
In particular, $\delta^{D(23)}_{LR}$, $\delta^{D(32)}_{LR} \lsim 
{\cal O}(10^{-2})$ is required in order to keep $BR(b \to s \gamma)$
within its experimental measured value.$^{[3]}$\footnotetext[3]{The 
bounds on 
the different delta's reported in \cite{misiak1,deltalimits} were obtained
using the mass insertion approximation, while we perform 
an exact diagonalization of the squark mass matrices. 
Therefore, in the cases where ${\cal O}(1)$ delta's are allowed (e.g., 
for $\delta^{D(32)}_{LL}$) these bounds may only serve as indicative  
for their expected size, since the mass insertion approximation necessarily
assumes small delta's.} On the other hand, 
$\delta^{D(23)}_{LL}$, $\delta^{D(32)}_{LL}$, $\delta^{D(23)}_{RR}$ and
$\delta^{D(32)}_{RR}$
of ${\cal O}(1)$ are not ruled out by 
$b \to s \gamma$ nor by any other low energy process that we know of. 

We will, thus, use the $LL$ and $RR$ delta's as the only source for 
$\tilde b - \tilde s$ mixing. 
Moreover, since there is no a-priory theoretical reason 
for the four $LL$ and $RR$ delta's to be significantly different,  
we will set all of them
to a common value denoted by $\delta^D$. That is, we fix
$\delta^{D(23)}_{LL}=\delta^{D(32)}_{LL}=\delta^{D(23)}_{RR}=
\delta^{D(32)}_{RR}=\delta^D$, and vary $\delta^D$
in the range $0 < \delta^D < 1$.       

\item{II.} The SUSY parameter space needed to
evaluate $\Gamma(Z\to b \bar s)$ in this scenario is:
$m_0,~\mu,~A,~\tan\beta,~m_{\tilde g}$ and $\delta^D$. 
The low-energy values of these six parameters  
fully determine 
the $\tilde b -\tilde s$ scalar spectrum 
(i.e., masses and mixing matrices) and the gluino mass ($m_{\tilde g}$), 
from which 
all the couplings in Table \ref{tab2} are calculated.
These six parameters will be varied subject to the requirement that 
squark masses as well as the gluino mass 
are heavier than 150 GeV. 
 
\end{description}

\begin{figure*}[htb]
\psfull
 \begin{center}
  \leavevmode
  \epsfig{file=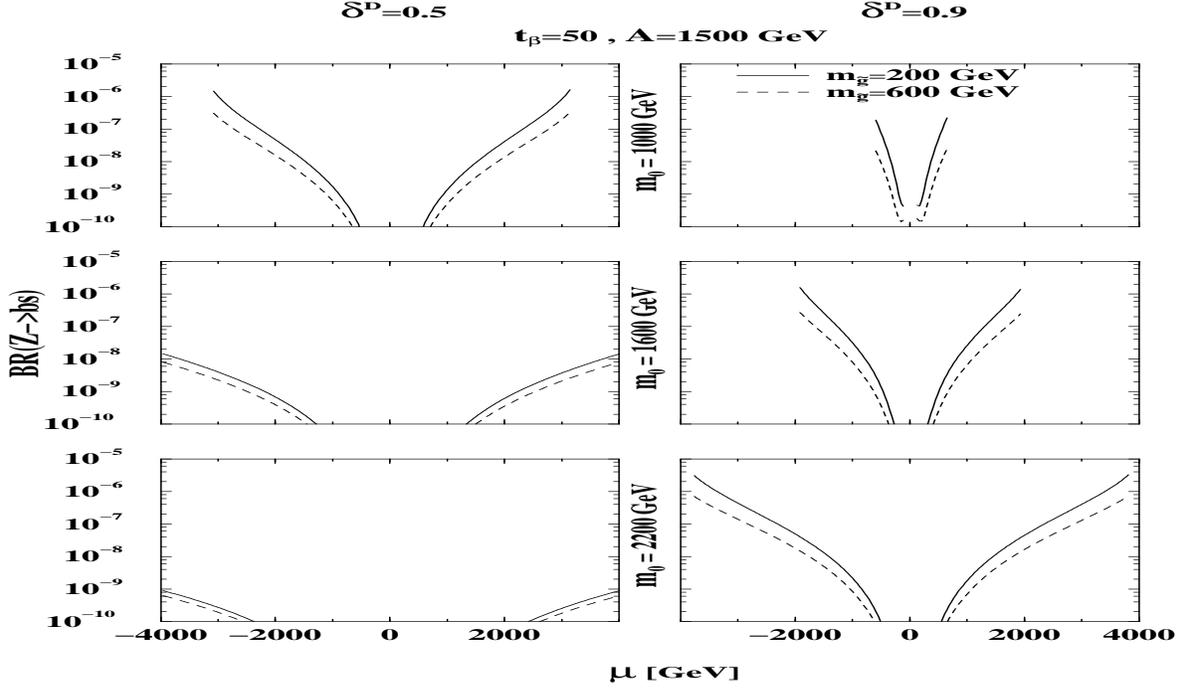,height=8cm,width=16cm,bbllx=0cm,bblly=2cm,bburx=20cm,bbury=25cm,angle=0}
 \end{center}
\caption{$BR(Z\to b \bar s + \bar b s)$ as a function 
of the Higgs mass parameter $\mu$, 
for combinations of $m_0=1000,~1600,~2200$ GeV with 
$m_{\tilde g}=200,~600$ GeV, for $t_\beta=50$, $A=1500$ GeV and 
for $\delta^D=0.5$ (left plots) and  
$\delta^D=0.9$ (right plots).}
\label{figbs1}
\end{figure*}

\begin{figure*}[htb]
\psfull
 \begin{center}
\vspace{1.cm}
  \leavevmode
  \epsfig{file=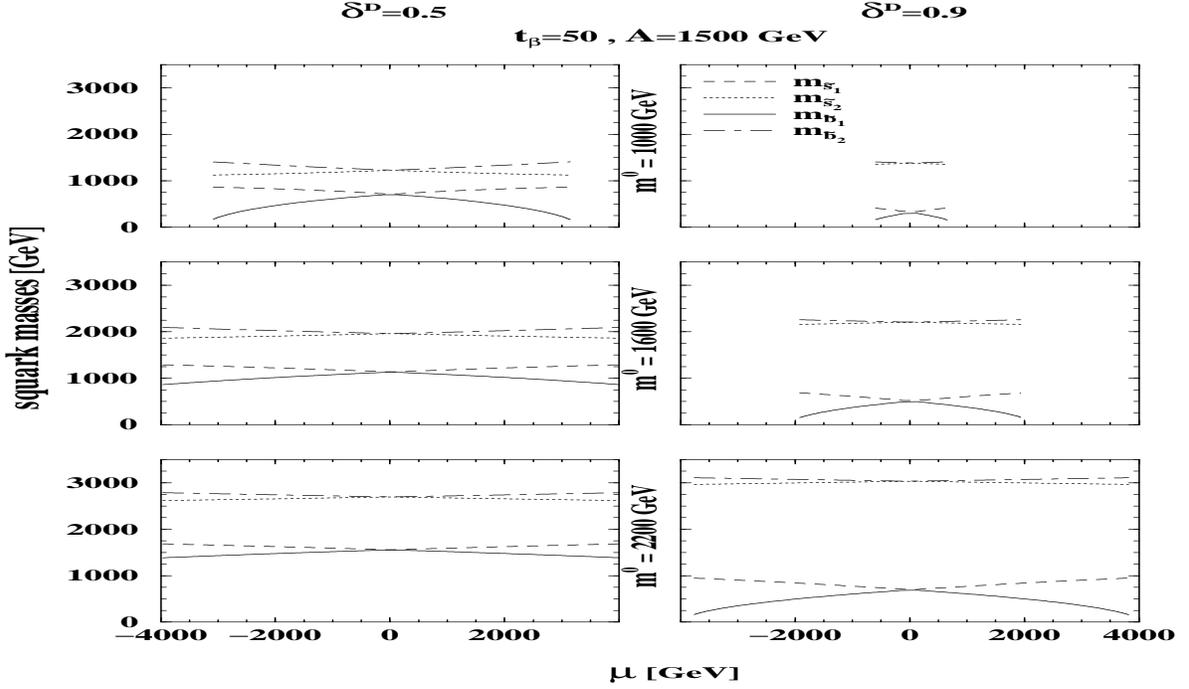,height=8cm,width=16cm,bbllx=0cm,bblly=2cm,bburx=20cm,bbury=25cm,angle=0}
 \end{center}
\caption{Physical masses of 
the second and third generation 
down-type squarks as a function $\mu$. The rest of the parameters 
are as in Fig.~\ref{figbs1}.}  
\label{figbs2}
\end{figure*}

\begin{figure*}[htb]
\psfull
 \begin{center}
  \leavevmode
  \epsfig{file=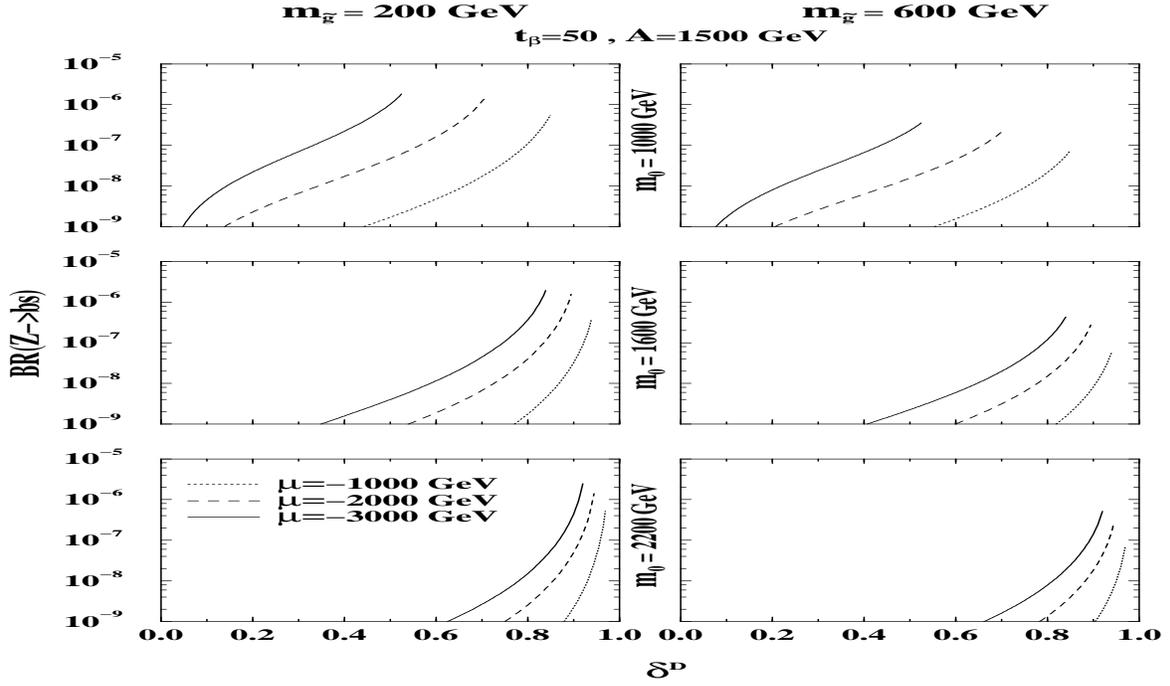,height=8cm,width=16cm,bbllx=0cm,bblly=2cm,bburx=20cm,bbury=25cm,angle=0}
 \end{center}
\caption{$BR(Z\to b \bar s + \bar b s)$ as a function 
of the flavor mixing parameter $\delta^D$, 
for combinations of $m_0=1000,~1600,~2200$ GeV with
$\mu=-1000,~-2000,~-3000$ GeV, for $t_\beta=50$, $A=1500$ GeV and 
for $m_{\tilde g}=200$ GeV (left plots) and  
$m_{\tilde g}=600$ GeV (right plots).}
\label{figbs3}
\end{figure*}

\begin{figure*}[htb]
\psfull
 \begin{center}
\vspace{1.cm}
  \leavevmode
  \epsfig{file=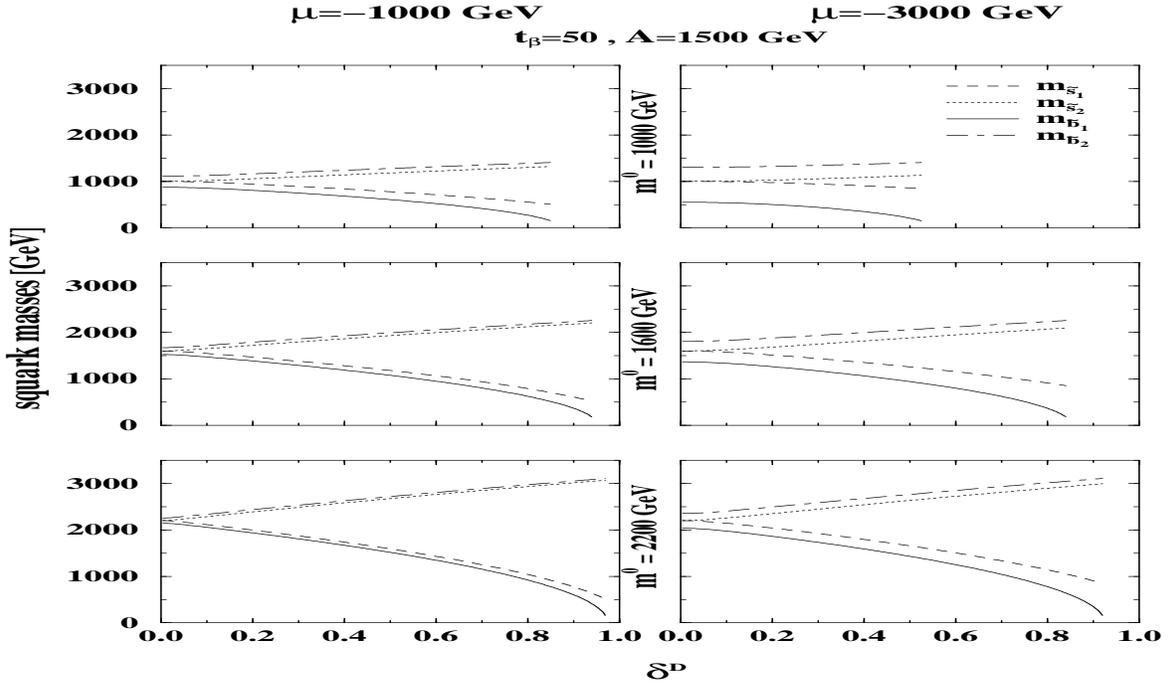,height=8cm,width=16cm,bbllx=0cm,bblly=2cm,bburx=20cm,bbury=25cm,angle=0}
 \end{center}
\caption{Physical masses of 
the second and third generation 
down-type squarks, as a function of $\delta^D$, 
for $m_0=1000,~1600,~2200$ GeV with
$\mu=-1000$ GeV (left plots) and
$\mu=-3000$ GeV (right plots).
The rest of the parameters 
are as in Fig.~\ref{figbs3}.}  
\label{figbs4}
\end{figure*}

\begin{figure*}[htb]
\psfull
 \begin{center}
  \leavevmode
  \epsfig{file=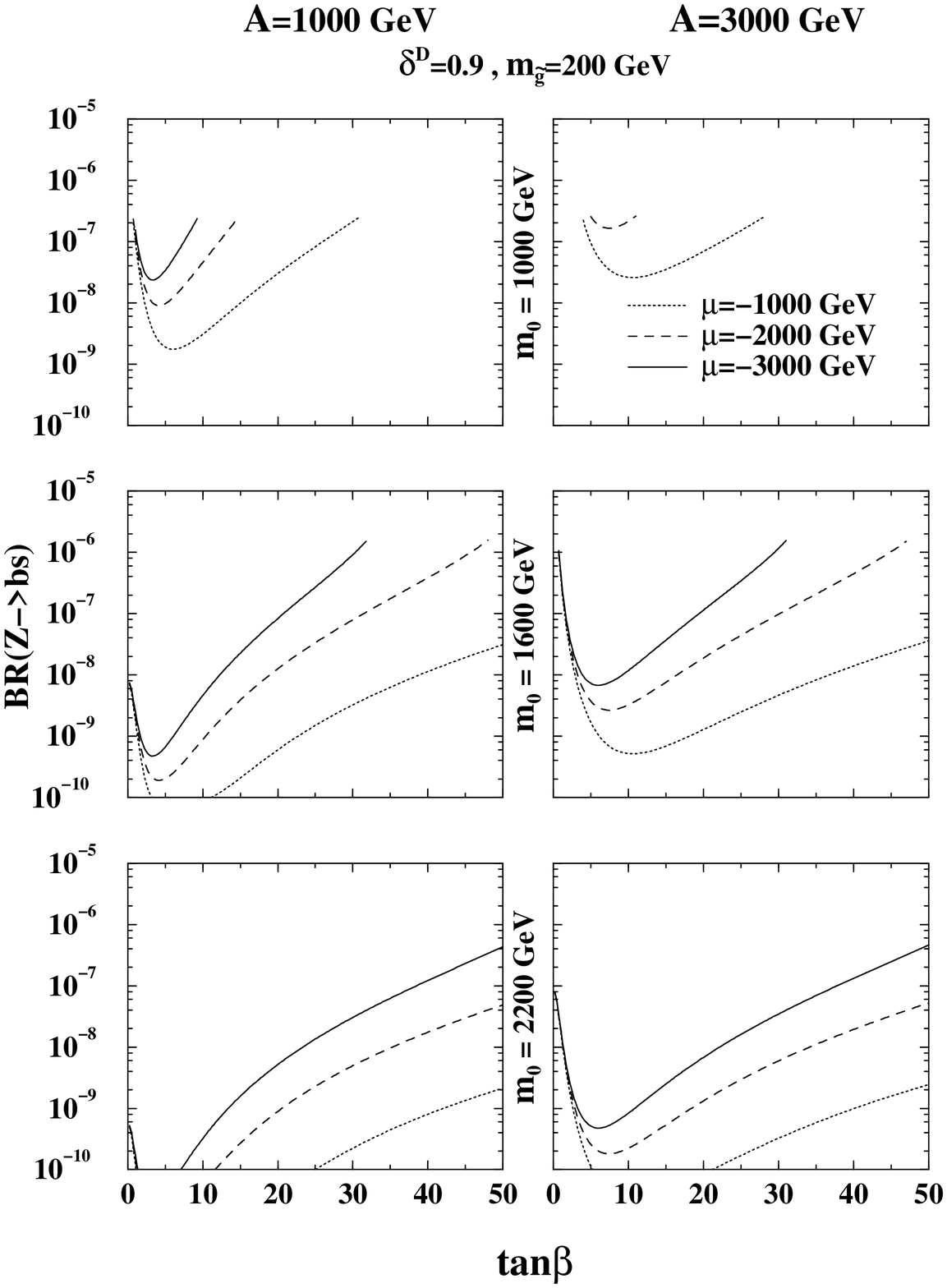,height=8cm,width=16cm,bbllx=0cm,bblly=2cm,bburx=20cm,bbury=25cm,angle=0}
 \end{center}
\caption{$BR(Z\to b \bar s + \bar b s)$ as a function 
of $\tan\beta$, 
for combinations of $m_0=1000,~1600,~2200$ GeV with
$\mu=-1000,~-2000,~-3000$ GeV, for $m_{\tilde g}=200$ GeV, $\delta^D=0.9$ GeV
and for $A=1000$ GeV (left plots) and  
$A=3000$ GeV (right plots).}
\label{figbs5}
\end{figure*}

In Figs.~\ref{figbs1}, \ref{figbs3} and \ref{figbs5} we 
plot $BR(Z\to b \bar s +\bar b s)$ as a function 
$\mu$, $\delta^D$ and $\tan\beta$, respectively, 
for three values of the common squark mass: $m_0 =1000, 1600$ or 2200 GeV. 
The rest of the parameters are varied subject to the above criteria 
\cite{curves}. 
In order to better understand the dependence of 
$BR(Z\to b \bar s +\bar b s)$ on the physical squark mass spectrum, 
we accompany Figs.~\ref{figbs1} and \ref{figbs3} by 
Figs.~\ref{figbs2} and \ref{figbs4}, respectively, in which we depict the 
masses of the four physical squarks 
$m_{\tilde s_{1,2}}$ and  $m_{\tilde b_{1,2}}$  
that correspond to the same choices of the SUSY parameter space as in 
Figs.~\ref{figbs1} and \ref{figbs3}.

%%%%%%%%%%%%%%%%%%%%%%%%%%%%%%%%%%%%%%%%%%%%%%%%%%%%%%%%%%%%%%%%%%
%\begin{eqnarray} 
%g_\gamma^{\alpha \beta} &=& \frac{1}{3}\delta_{\alpha \beta} ~.
%\end{eqnarray}
%
%g_\gamma^{\alpha \beta} &=& \frac{2}{3}\delta_{\alpha \beta} ~.
%\end{eqnarray}
%%%%%%%%%%%%%%%%%%%%%%%%%%%%%%%%%%%%%%%%%%%%%%%%%%%%%%%%%%%%%%%%%%

Let us summarize the results shown in Figs.~\ref{figbs1}-\ref{figbs5}:

\begin{itemize}

\item The branching ratio of the decay $Z \to b \bar s$ is 
enhanced dramatically 
with the increase of the mass splittings between the four physical squarks. 
This is due to a GIM-like cancellation which is operational 
in the limit 
of degenerate squark masses as a result of   
the unitarity of the rotation matrix $R_D$.$^{[4]}$\footnotetext[4]{The 
unitarity of 
$R_D$ also ensures that the infinities that appear in the individual 
diagrams of Fig.~\ref{fig1} cancel.}  
Thus for example, 
a typical mass spectrum that can drive the branching ratio to the 
$10^{-6}$ level is when the lightest down squark, $\tilde b_1$, has 
a mass below 250 GeV, while the rest of the squarks have masses 
at the 1-3 TeV range.

\item As expected, $BR(Z \to b \bar s)$ drops sharply 
as $\delta^D$ is decreased. Clearly, this is traced to the fact that 
the mixing among the bottom and strange type scalars diminishes in 
the limit $\delta^D \to 0$, see Figs.~\ref{figbs3} and \ref{figbs4}. 

\item For a sufficiently large $\tan\beta$, $BR(Z\to b \bar s)$ is almost 
insensitive to the value of the common trilinear soft breaking 
parameter $A$ as long as $\mu$ is large enough to drive the desired 
mass splittings between the squarks. 
This behavior is due to the dominance of the $\mu$ term 
in the quantity $X_b$ defined in (\ref{xb})
for large $t_\beta$ 
(recall that $X_b$ is responsible 
for the mass splitting between the two bottom-type scalars).
On the other hand, 
for $t_\beta \sim {\cal O}(1)$, the term $\propto A$ in $X_b$ 
(see the r.h.s. of (\ref{xb})) 
becomes important when $A \sim \mu$.
This feature can be seen in Fig.~\ref{figbs5}.

\item For the reason outlined above, 
$BR(Z\to b \bar s)$ is symmetric about $\mu=0$ for large 
$\tan\beta$, in which case the term $\propto \mu$ in $X_b$ dominates and 
the effect of the $A$ term is negligible.  

\item For $\mu^2/A^2 >> 1$, $BR(Z\to b \bar s)$ is increased with 
$\tan\beta$. Again, this is related to the 
dominance of the $\mu$ term in $X_b$
for large $\tan\beta$.

\item $BR(Z\to b \bar s)$ drops with $m_{\tilde g}$. 

\end{itemize}

To conclude this section, we have shown that 
$BR(Z\to b \bar s) \sim {\cal O}(10^{-6})$
can be achieved in the $\tilde b - \tilde s$ mixing scenario 
provided that the gluino mass and one of the third generation down-type 
scalar masses lie close to the electroweak mass scale, 
while the rest of the down-type squark masses
are at the TeV range, i.e., a large mass 
splitting between the lightest 
and rest of the down-type squarks is needed.
Such a mass hierarchy in the squark sector 
requires a typical ``heavy'' SUSY mass scale with soft breaking parameters 
at the level of a few TeV. This scenario
is somewhat motivated by the non-observability of SUSY particles 
in past and present high energy colliders.

\subsection{\bf $\tilde t - \tilde c$ mixing}

\begin{table*}[htb]
\begin{center}
%\begin{tabular}{||c|c||}
\begin{tabular}{c|c}
%\hline
%%%%%%%%%%%%%%%%%%%%%%%%%%%%%%%%%%%%%%%%%%%%%%%%%%%%
 & SUSY with $\tilde t - \tilde c$ mixing \\ 
~\\
\hline
~\\
scalar ($S_{\alpha}$) & $\Phi_{U,\alpha},~~\alpha=1,2,3,4$ \\
~\\
%\hline

fermion ($f_i$) & $\chi_i^c,~~i=1,2$ \\
~\\
%\hline
$a_{L(Zf)}^{ij}$ & $- \frac{e}{2s_Wc_W} a_{L(Z \chi^c)}^{ij}$  \\
~\\
%\hline
$a_{R(Zf)}^{ij}$ &  $- \frac{e}{2s_Wc_W} a_{R(Z \chi^c)}^{ij}$ \\
~\\
%\vspace{2.mm}%\hline
$b_{L(\alpha)}^{ij}$ & $f_L^{R(\alpha i j)\star} 
\left( R_{U,1\alpha}^{\star} V^{2j}_{CKM} +  
R_{U,3\alpha}^{\star} V^{3j}_{CKM} \right)
+ \frac{f_R^{R(\alpha i j)\star}}{m_{u_\alpha}}
\left( m_{c} R_{U,2\alpha}^{\star} V^{2j}_{CKM} +  
m_{t} R_{U,4\alpha}^{\star} 
V^{3j}_{CKM} \right) $ \\
~\\
%\vspace{2.mm}
%\hline
$b_{R(\alpha)}^{ij}$ & $f_L^{L(\alpha i j)\star} 
\left( R_{U,1\alpha}^{\star} V^{2j}_{CKM} +  
R_{U,3\alpha}^{\star} V^{3j}_{CKM} \right)$ \\
~\\
%\hline
$g_Z^{\alpha \beta}$ & $\frac{e}{2s_Wc_W} 
\left( R_{U,1 \alpha}^{\star} R_{U,1 \beta}
+  R_{U,3 \alpha}^{\star} R_{U,3 \beta} - 
\frac{4}{3} s_W^2 \delta_{\alpha \beta} 
\right)$ \\
~\\
%\hline
\end{tabular} 
\caption{The couplings required for the calculation 
of $\Gamma(Z\to b \bar s)$ in the $\tilde t - \tilde c$ mixing scenario.
The couplings follow the notation in (\ref{Vff}), (\ref{VSS}) and 
(\ref{Sdf}).
Also, $a_{L(Z \chi^c)}^{ij}$ and $a_{R(Z \chi^c)}^{ij}$
are given in (\ref{vcc1}) and (\ref{vcc}), 
$f_L^{R(\alpha i j)}$, $f_R^{R(\alpha i j)}$ and
$f_L^{L(\alpha i j)}$ are defined in (\ref{fRfL}). 
$R_U$ is the rotation matrix defined in (\ref{rotate}).}
\label{tab3}
\end{center}
\end{table*}

In the stop-scharm mixing scenario the 
flavor changing decay $Z\to b \bar s$ proceeds through 
one-loop exchanges of the 
$\tilde t - \tilde c$ admixture states, $\Phi_U$, and 
charginos, $\chi$. More specifically, 
we have $S_\alpha = \Phi_{U,\alpha}$ 
with $\alpha=1-4$,
and $f_i = \chi^c_i$ with $i=1,2$ for the two charginos 
(we find it convenient to 
calculate the exchanges of the charged conjugate chargino states $\chi^c$).

\begin{figure*}[htb]
\psfull
 \begin{center}
\vspace{1.cm}
  \leavevmode
  \epsfig{file=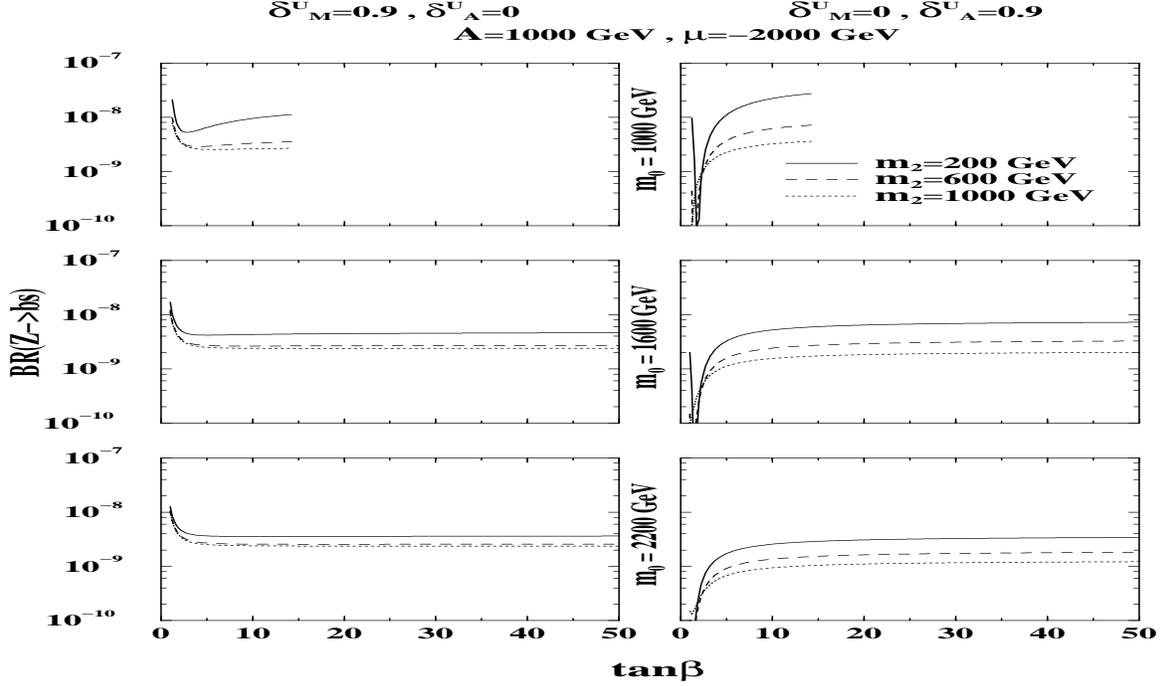,height=8cm,width=16cm,bbllx=0cm,bblly=2cm,bburx=20cm,bbury=25cm,angle=0}
 \end{center}
\caption{$BR(Z\to b \bar s + \bar b s)$ as a function 
of $\tan\beta$, 
for combinations of $m_0=1000,~1600,~2200$ GeV with 
$m_2=200,~600,~1000$ GeV, for $A=1000$ GeV, $\mu=-2000$ GeV and 
for $\delta^U_M=0.9,~\delta^U_A=0$ (left plots) and  
$\delta^U_M=0,~\delta^U_A=0.9$ (right plots).}
\label{figtc1}
\end{figure*}

\begin{figure*}[htb]
\psfull
 \begin{center}
  \leavevmode
  \epsfig{file=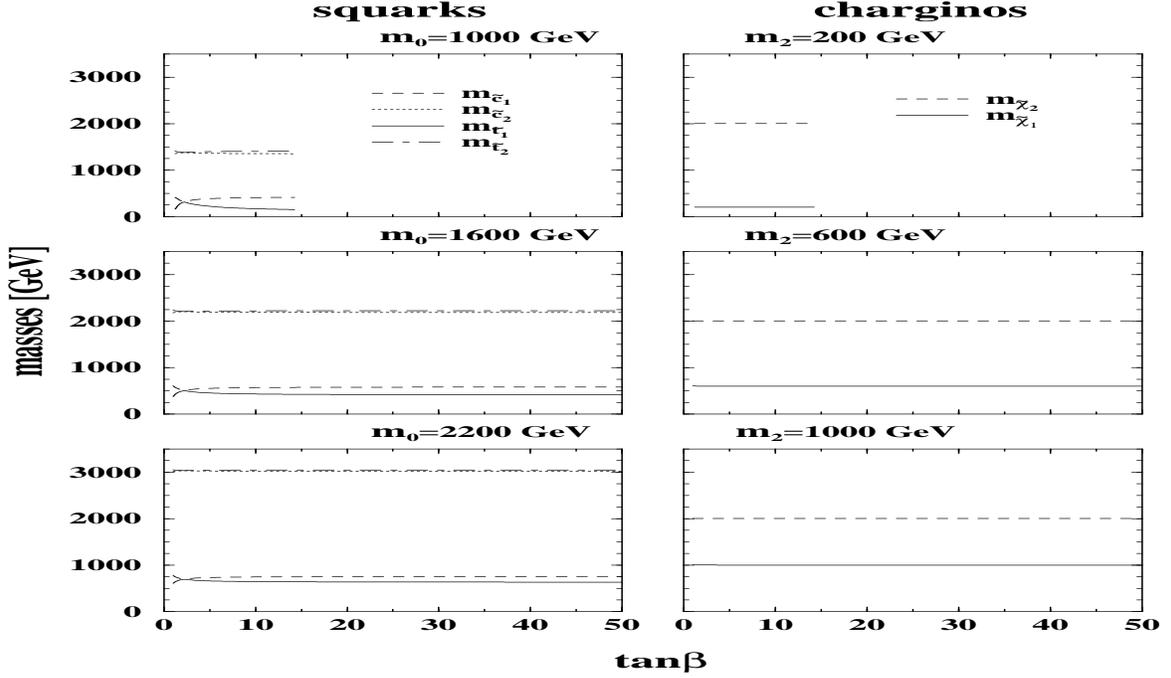,height=8cm,width=16cm,bbllx=0cm,bblly=2cm,bburx=20cm,bbury=25cm,angle=0}
 \end{center}
\caption{Physical masses of 
the second and third generation 
up-type squarks (left plots)  
and of the charginos (right plots), 
as a function of $\tan\beta$. Squark masses are shown for 
$m_0=1000,~1600$ and $2200$ GeV and for $\delta^U_M=0.9,~\delta^U_A=0$ 
(equivalent to the case $\delta^U_M=0,~\delta^U_A=0.9$, see text) and  
chargino masses are given 
for $m_2=200,~600$ and $1000$ GeV.  
The rest of the parameters 
are as in Fig.~\ref{figtc1}.}  
\label{figtc2}
\end{figure*}

Thus, in the $\tilde t - \tilde c$ mixing scenario 
the following interaction vertices are required:

\begin{eqnarray}
V_\mu f_i \bar f_j &\to& Z_\mu \chi_i^c \bar\chi_j^c ~,\nonumber \\
V_\mu S_\alpha S_\beta &\to& Z_\mu \Phi_{U,\alpha}^\star \Phi_{U,\beta} ~, \\
S_\alpha \bar f_i d_j &\to& \Phi_{U,\alpha}^\star \bar\chi^c_i d_j ~.\nonumber
\end{eqnarray}

These vertices are taken from \cite{rosiek}:

\begin{widetext}
\begin{eqnarray}
{\cal L}(V_\mu \chi_i^c \bar\chi_j^c) &=&
- \frac{e}{2 s_Wc_W} \bar\chi_j^c \gamma_\mu 
\left(a_{L(V \chi^c)}^{ij}L +
 a_{R(V \chi^c)}^{ij}R \right) \chi_i^c V_\mu 
\label{frtc1}~,\\
{\cal L}(V_\mu \tilde u \tilde u) &=& -i
\left[ \frac{1}{2} \frac{e}{s_W}
 A_\mu^3 + \frac{1}{6} \frac{e}{c_W} B_\mu \right] 
 \tilde u_{L,\ell}^{\star} \stackrel{\leftrightarrow}{\partial^\mu_{\tilde u}} 
\tilde u_{L,\ell}
+ i
 \frac{2}{3} \frac{e}{c_W}  B_\mu \tilde u_{R,\ell}
\stackrel{\leftrightarrow}{\partial^\mu_{\tilde u}} \tilde u_{R,\ell}^{\star} 
\label{vulul}~,\\
{\cal L}(\tilde u \chi^c d) &=& 
\tilde u_{L,\ell} \bar d_j \left( f_L^{L(\ell i j)} L + 
f_L^{R(\ell i j)} R \right) \chi_i^c V^{\ell j \star}_{CKM}
+ \tilde u_{R,\ell} \bar d_j \left( f_R^{L(\ell i j)} L + 
f_R^{R(\ell i j)} R \right) \chi_i^c V^{\ell j \star}_{CKM} +h.c.
\label{sdfsusy}~,
\end{eqnarray}
\end{widetext}

\noindent where 

\begin{eqnarray}
 a_{L(Z \chi^c)}^{ij} &=&  \left( Z_{1 i}^- Z_{1 j}^{- \star} + 
\cos 2 \theta_W \delta_{ij} \right) \label{vcc1}~, \\
 a_{R(Z \chi^c)}^{ij} &=& \left( Z_{1 i}^{+ \star} Z_{1 j}^{+} + 
\cos 2 \theta_W \delta_{ij} \right) \label{vcc}~,
\end{eqnarray}

\begin{eqnarray}
f_L^{L(\ell i j)} &=& \frac{e}{\sqrt 2 s_W} 
\frac{m_{d_j}}{M_W \cos\beta} Z_{2i}^- ~, \nonumber \\
f_L^{R(\ell i j)} &=& -\frac{e}{s_W} Z_{1i}^{+ \star} ~, \nonumber \\
f_R^{L(\ell i j)} &=& 0 ~,\nonumber \\
f_R^{R(\ell i j)} &=& \frac{e}{\sqrt 2 s_W} 
\frac{m_{u_\ell}}{M_W \sin\beta}
Z_{2i}^{+ \star} \label{fRfL}~.
\end{eqnarray}

%%%%%%%%%%%%%%%%%%%%%%%%%%%%%%%%%%%%%%%%%%%%%%%%%%%%%%%%%%%%%%%
% a_{L(\gamma \chi^c)}^{ij} &=&  a_{R(\gamma \chi^c)}^{ij} = 
%- e \delta_{ij} ~,\nonumber
%%%%%%%%%%%%%%%%%%%%%%%%%%%%%%%%%%%%%%%%%%%%%%%%%%%%%%%%%%%%%%%

\noindent and $Z^\pm$ are the chargino mixing matrices given 
in \cite{rosiek}. Also, $A^3$ and $B$ are the SU(2) and U(1) gauge fields, 
respectively, and  $\tilde u_{L,R}$ are
the SU(2) weak states of the up-type scalars.

Here also, the couplings 
needed for the calculation of $\Gamma(Z \to b \bar s)$ in the form 
defined in (\ref{Vff}), (\ref{VSS}) and (\ref{Sdf}) are obtained from 
the Lagrangian terms in 
(\ref{frtc1})-(\ref{fRfL})
by 
rotating the weak states,  
$\Phi^0_{U}$, to the physical 
states, $\Phi_{U}$, according to (\ref{rotate}). These couplings  
are summarized in Table \ref{tab3}.

The contribution of the $\tilde t -\tilde c$ mixed states to the 
one-loop diagrams in Fig.~\ref{fig1} are characterized as follows:  

\begin{description}

\item{I.} The quantities that mediate the 
flavor changing transition $b \to s$ in the 
$\tilde t -\tilde c$ mixed scenario are:
$\delta^{U(23)}_{LL}$, $\delta^{U(32)}_{LL}$, $\delta^{U(23)}_{RR}$,
$\delta^{U(32)}_{RR}$, $\delta^{U(23)}_{LR}$ and $\delta^{U(32)}_{LR}$. 
Recall that the $LL$ and $RR$ delta's originate 
from the bilinear soft terms in (\ref{soft}), while 
the $LR$ delta's are associated with the trilinear soft breaking 
SUSY terms.
Thus, we will 
separate these two types of flavor violating sources 
in our numerical analysis.
In particular, we define 
$\delta^U_M=\delta^{U(23)}_{LL}=\delta^{U(32)}_{LL}=\delta^{U(23)}_{RR}=
\delta^{U(32)}_{RR}$ and $\delta^U_A=\delta^{U(23)}_{LR}=\delta^{U(32)}_{LR}$
and we vary either $\delta^U_M$ or $\delta^U_A$ in the range 
$\left[0,1 \right]$. Note that  
an ${\cal O}(1)$ value for either $\delta^U_M$ or $\delta^U_A$ is consistent 
with all experimental data \cite{misiak1,deltalimits}. 

\item{II.} The required SUSY parameter space is:
$m_0,~\mu,~A,~\tan\beta,~m_2$ and $\delta^U_M,~\delta^U_A$, where 
$m_2$ is the SU(2) gaugino mass parameter.  
The low-energy values of these six parameters  
fix  
the $\tilde t -\tilde c$ scalar spectrum 
(i.e., masses and mixing matrices) and the chargino masses 
and mixing matrices from which 
all couplings in Table \ref{tab3} are derived.

As in the $\tilde b -\tilde s$ mixing case, 
these parameters will be varied subject to the requirement that 
the squark masses are heavier than 150 GeV and, in addition, that 
the charginos are heavier than 100 GeV \cite{curves}.
 
\end{description}

Taking maximal flavor violation in the $\tilde t - \tilde c$ mixing 
scenario, i.e.,   
$\delta^U_M \sim {\cal O}(1)$ or
$\delta^U_A \sim {\cal O}(1)$, and varying the rest of the SUSY 
parameters involved subject to the above criteria, we find that 
$BR(Z\to b \bar s +\bar b s)$ can reach ${\rm few} \times 10^{-8}$ 
at best. Here also, the 
$BR(Z\to b \bar s +\bar b s)$ is significantly enhanced when large 
mass splittings between the four up-type squarks, 
$m_{\tilde c_{1,2}}$ and  $m_{\tilde t_{1,2}}$ are present. 
Such a hierarchy in the up-type squark mass spectrum 
makes the GIM-like cancellation mentioned earlier less effective. 

Indeed a two orders of magnitude difference between the $\tilde t - \tilde c$ 
and $\tilde b - \tilde s$ mixing cases is expected due to 
an $(\alpha_s/\alpha)^2$ enhancement factor 
in the $\tilde b - \tilde s$ scenario (compared to the $\tilde t - \tilde c$
mixing case) 
which arises from the gluino QCD coupling. 

In Figs.~\ref{figtc1} we 
plot $BR(Z\to b \bar s +\bar b s)$ as a function 
of $\tan\beta$, 
for combinations of $m_0 =1000, 1600$ and 2200 GeV with
$m_2 =200, 600$ and 1000 GeV and for either 
$\left\{\delta^U_M=0.9,~\delta^U_A=0\right\}$ or 
$\left\{\delta^U_M=0,~\delta^U_A=0.9\right\}$.$^{[5]}$\footnotetext[5]{Note 
that 
the physical up-squark masses have the same dependence 
on $\delta^U_M$ or $\delta^U_A$ when one of the two delta's is set to zero. 
Thus, for example 
$m_{\tilde u_i}(\delta^U_M=0.9,\delta^U_A=0)=
m_{\tilde u_i}(\delta^U_M=0,\delta^U_A=0.9)$, $i=1,2,3$ and 4.} For illustration we set
$A=1000$ GeV and $\mu=-2000$ GeV.  
In Figs.~\ref{figtc2} we depict the 
masses of the four physical up-type squarks 
$m_{\tilde c_{1,2}}$ and  $m_{\tilde t_{1,2}}$ and 
the masses of the two chargino states as a function of $\tan\beta$,    
for the same SUSY parameter choices as in 
Figs.~\ref{figtc1} \cite{curves}.

\section{SUSY with $R_P$ violating interactions}

If $R_P$ is violated in the SUSY superpotential, then 
flavor changing transitions can emerge from interactions of squarks 
or sleptons with fermions.
In particular, there are two types of RPV terms that are allowed in 
the superpotential if the discrete $R_P$ symmetry is not imposed. 
These are the RPV trilinear Yukawa-like (RPVT) operators 
and bilinear (RPVB) operators.

In the usual convention, the RPVT are proportional to the dimensionless 
couplings  
$\lambda,~\lambda^\prime$ and $\lambda^{\prime \prime}$,
see e.g., \cite{rpvreview}. 
Here we will assume that $\lambda^{\prime \prime} \ll \lambda^\prime$ 
and investigate the one-loop 
effects of the $\lambda^\prime$ type operator
on our flavor changing decay $Z \to d_I \bar d_J$:$^{[6]}$\footnotetext[6]{Note
 that at the one-loop 
level the $\lambda$ type couplings do not contribute to the decay 
$Z \to d_I \bar d_J$.}    

\begin{eqnarray}
{\cal W}_{RPVT} \supset \epsilon_{ab}
\lambda_{ijk}^\prime {\hat L}^a_i {\hat Q}^b_j {\hat D}_k^c
 \label{rpvt}~,
\end{eqnarray}

\noindent where ${\hat Q}$ and ${\hat L}$ are SU(2) doublet quark 
and lepton supermultiplets, respectively, and ${\hat D}$ is the 
SU(2) singlet down-type quark supermultiplet. 
Also, $i,j,k=1,2$ or 3 are generation indices and $a,b$ are SU(2) indices.

The RPVB operator is:

\begin{eqnarray}
{\cal W}_{RPVB} = - \epsilon_{ab} \mu_i 
{\hat L}^a_i \hat H_u^b 
\label{rpvb}~,
\end{eqnarray}

\noindent where $\hat H_u$ is the up-type Higgs supermultiplet and 
$i=1,2$ or 3 labels the lepton generation.

In addition, if one does not impose $R_P$, then the
usual set of $R_P$ conserving (RPC) soft SUSY breaking terms 
is extended by new
trilinear and bilinear soft terms which correspond to the RPV terms
of the superpotential, i.e., to the ones in (\ref{rpvt}) and (\ref{rpvb}). 
For our
purpose, 
only the following soft SUSY breaking bilinear term is 
relevant \cite{GH,morebterms,davidson}:

\begin{equation}
V_{RPVB} = \epsilon_{ab} b_i {\tilde L}^a_i H_u^b \label{bterm}~,
\end{equation}

\noindent where $\tilde L$ and $H_u$ are the scalar components of
$\hat L$ and $\hat H_u$, respectively.

The RPVT operator ($\propto \lambda^\prime$) in (\ref{rpvt}) 
gives rise to the following scalar-fermion-fermion RPV interactions:

\begin{eqnarray}
{\cal L} = \lambda_{ijk}^\prime \left\{
\tilde\nu_L^i \bar d^k_R d_L^j + 
\tilde d_L^j  \bar d^k_R \nu_L^i +
\left(\tilde d_R^k \right)^* \left(\bar{\nu}_L^{i} \right)^c d_L^j 
\right. \nonumber\\
\left. - \tilde e_L^i  \bar d^k_R u_L^j - 
\tilde u_L^j  \bar d^k_R e_L^i -
\left(\tilde d_R^k \right)^* \left(\bar e_L^{i} \right)^c u_L^j \right\} +h.c.
\label{rpv} ~.
\end{eqnarray}

\noindent where $d(u)$ is a down(up)-quark, $e(\nu)$ is a 
charged-lepton(neutrino) and scalars are denoted with a tilde.

The RPVB operator ($\propto \mu_i$) in (\ref{rpvb}) gives rise to mixings
among charged leptons and charginos as well as between neutrinos and 
neutralinos. However,   
low energy flavor changing processes \cite{ref12}, 
flavor changing leptonic Z-decays \cite{ref8} and neutrino 
masses \cite{GH,ref8,neutrinomass,basis} suggest that 
the $\mu_i$ are expected to be vanishingly small.  
We will, therefore, neglect its 
contribution to the decay $Z \to d_I \bar d_J$.$^{[7]}$\footnotetext[7]{The 
one-loop exchanges of possible lepton-chargino and neutrino-neutralino 
admixture states in $Z \to d_I \bar d_J$ will be controlled by the 
square of the RPV couplings product 
$\mu_i \times \lambda^\prime$.}
On the other hand, the soft breaking RPVB term ($\propto b_i$) in 
(\ref{bterm}) gives rise to mixings between sleptons and Higgs-bosons 
which may be exchanged in the loops of the diagrams shown in Fig.\ref{fig1}.

Let us now categorize the different types of RPV interactions that
contribute at one-loop to the 
flavor changing decay of $Z\to d_I \bar d_J$. 
Since the decay $Z\to d_I \bar d_J$ conserves $R_P$, there should 
be two insertions of RPV vertices in the one-loop diagrams of Fig.\ref{fig1}.
We can thus divide the various types of RPV one-loop 
exchanges into two categories, type A and type B,  
according to the pair of 
RPV couplings involved:

\begin{description}

\item{\bf Type A:} The RPV contributions that are proportional to the product 
$\lambda^\prime \lambda^\prime$, i.e., $\Gamma(Z\to d_I \bar d_J) \propto 
\lambda^\prime \lambda^\prime$, where $\lambda^\prime$ is defined in 
(\ref{rpvt}).

\item{\bf Type B:} The RPV contributions that are proportional to the product 
$b \lambda^\prime$, i.e., $\Gamma(Z\to d_I \bar d_J) \propto 
b \lambda^\prime$, where $b$ is the soft breaking 
RPV bilinear coupling defined in (\ref{bterm}). 

\end{description}

\subsection{Type A RPV effect}

The type A RPV contribution to 
$Z \to d_I \bar d_J$  
emanates from 
the first five RPV Yukawa-like interaction vertices in (\ref{rpv}).
In this case we assume that $b_i \to 0$ such that 
mixing effects between sleptons and the Higgs fields are absent.

We can further sub-divide the type A contributions into
6 types according to the type of scalar ($S$) and type of fermion ($f$)
that are being exchanged in the loops:

\begin{eqnarray}
{\rm type~A1:} && S_\alpha=\tilde e_{L,\alpha}~~,~~f_i=u_i \nonumber \\
{\rm type~A2:} && S_\alpha=\tilde d_{L,\alpha}~~,~~f_i=\nu_i \nonumber \\
{\rm type~A3:} && S_\alpha=\tilde d_{R,\alpha}~~,~~f_i=\nu^c_i \nonumber \\
{\rm type~A4:} && S_\alpha=\tilde\nu_{L,\alpha}~~,~~f_i=d_i \nonumber \\
{\rm type~A5:} && S_\alpha=\tilde\nu_{L,\alpha}^*~~,~~f_i=d_i \nonumber \\
{\rm type~A6:} && S_\alpha=\tilde u_{L,\alpha}~~,~~f_i=e_i ~, 
\end{eqnarray}

\n where $\alpha=1,2,3$ and $i=1,2,3$.

\begin{table*}[htb]
\begin{center}
%\begin{tabular}{||c|c|c|c||}
\begin{tabular}{c|c|c|c}
%\hline
%%%%%%%%%%%%%%%%%%%%%%%%%%%%%%%%%%%%%%%%%%%%%%%%%%%%
 & type A1 & type A2 & type A3 \\ 
~ & ~ & ~ & \\
\hline
~ & ~ & ~ & \\
scalar ($S_{\alpha}$) & 
$\tilde e_{L,\alpha},~~\alpha=1,2$   & 
$\tilde d_{L,\alpha},~~\alpha=1,2,3$ & 
$\tilde d_{R,\alpha},~~\alpha=1,2,3$ 
\\ 
~ & ~ & ~ & \\
%\hline
fermion ($f_i$) & 
$u_i,~~i=1,2,3$ &  
$\nu_i,~~i=1,2,3$ &
$\nu_i^c,~~i=1,2,3$ 
\\ 
~ & ~ & ~ & \\
%\hline
$a_{L(Zf)}^{ij}$ & 
$a_{L(Zu)}$& 
$a_{L(Z \nu)}$ & 
$-a_{R(Z \nu)}$  
\\
~ & ~ & ~ & \\
%\hline
$a_{R(Zf)}^{ij}$ & 
$a_{R(Zu)}$& 
$a_{R(Z \nu)}$ & 
$-a_{L(Z \nu)}$  
\\
~ & ~ & ~ & \\ 
%\hline
$b_{L(\alpha)}^{ij}$ & 
$0$ &
$0$ &
$\lambda^{\prime}_{i j \alpha}$ 
\\
~ & ~ & ~ & \\ 
%\hline
$b_{R(\alpha)}^{ij}$ & 
$- \lambda^{\prime *}_{\alpha i j}$ &
$\lambda^{\prime *}_{i \alpha j}$ &
$0$ 
\\
~ & ~ & ~ & \\ 
%\hline
$g_Z^{\alpha \beta}$ &
$- e \frac{c_W^2-s_W^2}{2s_Wc_W} \delta_{\alpha \beta}$ &
$- \frac{e}{2s_Wc_W } \left(1 - \frac{2}{3} s_W^2 \right) 
\delta_{\alpha \beta}$ &
$\frac{1}{3} e \frac{s_W}{c_W}  \delta_{\alpha \beta}$ 
\\
~ & ~ & ~ & \\ 
%\hline
\end{tabular}

%\vspace{0.5cm}
\medskip

%\begin{tabular}{||c|c|c|c||}
\begin{tabular}{c|c|c|c}
%\hline
%~ & ~ & ~ & \\
%%%%%%%%%%%%%%%%%%%%%%%%%%%%%%%%%%%%%%%%%%%%%%%%%%%%
 & type A4 & type A5 & type A6
\\ 
~ & ~ & ~ & \\
\hline
~ & ~ & ~ & \\
scalar ($S_{\alpha}$) & 
$\tilde\nu_{L,\alpha},~~\alpha=1,2$ &
$\tilde\nu_{L,\alpha}^\star,~~\alpha=1,2$ &
$\tilde u_{L,\alpha},~~\alpha=1,2,3$
\\
~ & ~ & ~ & \\ 
%\hline
fermion ($f_i$) & 
$d_i,~~i=1,2,3$ &
$d_i,~~i=1,2,3$ &
$e_i,~~i=1,2,3$ 
\\ 
~ & ~ & ~ & \\
%\hline
$a_{L(Zf)}^{ij}$ & 
$a_{L(Z d)}$ & 
$a_{L(Z d)}$ & 
$a_{L(Z e)}$ 
\\
~ & ~ & ~ & \\
%\hline
$a_{R(Zf)}^{ij}$ & 
$a_{R(Z d)}$ & 
$a_{R(Z d)}$ & 
$a_{R(Z e)}$ 
\\ 
~ & ~ & ~ & \\
%\hline
$b_{L(\alpha)}^{ij}$ & 
$\lambda^{\prime}_{\alpha j i}$ &
$0$ &
$0$ 
\\
~ & ~ & ~ & \\ 
%\hline
$b_{R(\alpha)}^{ij}$ & 
$0$ &
$\lambda^{\prime * }_{\alpha i j}$ &
$- \lambda^{\prime *}_{i \alpha j}$
\\
~ & ~ & ~ & \\ 
%\hline
$g_Z^{\alpha \beta}$ &
$- \frac{e}{2s_Wc_W} \delta_{\alpha \beta}$ &
$\frac{e}{2s_Wc_W} \delta_{\alpha \beta}$ &
$\frac{e}{2s_Wc_W} \left( 1 - \frac{4}{3}s_W^2 \right) 
\delta_{\alpha \beta}$
\\
~ & ~ & ~ & \\ 
%\hline
\end{tabular} 
\caption{The couplings required for the calculation 
of $\Gamma(Z\to d_I \bar d_J)$ in the type A RPV scenario.
The couplings follow the notation in 
(\ref{Vff})-(\ref{Sdf}). 
\label{tabtypeA}}
\end{center}
\end{table*}

For each of the type A RPV exchanges above, 
the generic couplings defined in (\ref{Vff}), (\ref{VSS}) and (\ref{Sdf}) 
are summarized in Table \ref{tabtypeA}.   
In particular, for a given $f$, 
the $Zff$ couplings of (\ref{Vff}) are given by 
(\ref{Vdd})-(\ref{Vddlast}). The $Sdf$ couplings are taken from 
the Yukawa like interactions in (\ref{rpv}), while  
the $ZSS$ couplings are 
extracted from ${\cal L}(V_\mu \tilde u \tilde u)$ in 
(\ref{vulul}), from ${\cal L}(V_\mu \tilde d \tilde d)$
in (\ref{vdldl}) and from the $V \tilde L \tilde L$ interaction 
Lagrangian: 

\begin{equation}
{\cal L}(V_\mu \tilde L \tilde L) = -i \frac{1}{2}
 \tilde L_{\ell}^{\star} \left[  \frac{e}{s_W}
 \tau^3 A_\mu^3 -  \frac{e}{c_W} B_\mu \right] 
 \stackrel{\leftrightarrow}{\partial^\mu_{\tilde L}} 
\tilde L_{\ell} \label{velel}~,
\end{equation}

\noindent where 
$A^3$ and $B$ are the SU(2) and U(1) gauge fields, 
respectively, $\tilde L = \pmatrix{
\tilde\nu_L \cr \tilde e_L }$ and 
$\tau^3=\pmatrix{
1 & 0 \cr
0 & -1 }
$.

%%the couplings of a photon to a pair of scalars:
%
%type 1: g_\gamma^{\alpha \beta} &=& -e \delta_{\alpha \beta} ~,
%type 2: g_\gamma^{\alpha \beta} &=& -\frac{e}{3} \delta_{\alpha \beta}~,
%type 3: g_\gamma^{\alpha \beta} &=& -\frac{e}{3} \delta_{\alpha \beta}  ~,
%type 4: g_\gamma^{\alpha \beta} &=& 0 ~,
%type 5: g_\gamma^{\alpha \beta} &=& 0 ~,
%type 6: g_\gamma^{\alpha \beta} &=& \frac{2}{3} e \delta_{\alpha \beta} ~,

Given the couplings in Table \ref{tabtypeA} and 
the structure of the form factors in (\ref{firstamp})-(\ref{lastamp}) 
it is evident 
that there are only two types of $\lambda^\prime \lambda^\prime$ product
combinations which enter the 
type A RPV contribution to the decay $Z \to d_I \bar d_J$:

\begin{enumerate}

\item The product $\lambda^{\prime}_{m n I}  
\lambda^{\prime *}_{m n J}$. 
Types A1, A2, A5 and A6 are proportional to this couplings product.

\item The product $\lambda^{\prime *}_{m I n}  
\lambda^{\prime}_{m J n}$.
Types A3 and A4 are proportional to this couplings product. 

\end{enumerate}

Furthermore, since none of the scalars have 
both a left and a right handed RPVT coupling to fermions 
in the type A scenario, 
i.e., in the notation of (\ref{Sdf}) either $b_{L(\alpha)}^{ij}=0$
or  $b_{R(\alpha)}^{ij}=0$ (see Table \ref{tabtypeA}), the form factors 
$B_{L,k}^{IJ}$ and $B_{R,k}^{IJ}$ in the amplitude (\ref{muk})
(which requires a none-zero value for both the 
left and the right handed scalar-fermion-down quark 
couplings) vanish. 
Also, since $b_{L(\alpha)}^{ij}=0$ for the RPV 
contributions of types A1, A2, A5 and A6, they contribute 
only to the right-handed vector-like form factor $A_{R,k}^{IJ}$. 
Similarly, the RPV contributions of types A3 and A4 
have $b_{R(\alpha)}^{ij}=0$, therefore, contributing only 
to $A_{L,k}^{IJ}$.   
 
It should be noted that for any one of the type A RPV exchanges,  
if the scalars of different flavors that are being exchanged in 
the loops are degenerate and upon neglecting all fermion masses 
except for the top-quark, then there remain only 
three distinct types of contributions 
of the $\lambda^\prime$ products 
in the type A RPV scenario. That is, 
under this assumption 
$BR(Z \to d_I \bar d_J)$ can have only three different values which 
we denote by $BR1^{IJ},~BR2^{IJ}$ and $BR3^{IJ}$ as follows:

\begin{widetext}
\begin{eqnarray}
BR1^{IJ}  
& = & BR(Z \to d_I \bar d_J) ~ {\rm when}~ 
\left( \lambda^\prime_{ijI} \times \lambda^\prime_{ijJ} \right)^2 \neq 0 
~;~ j \ne 3 ,~ i=1,2,3 \label{br1} \\
BR2^{IJ} 
& = & BR(Z \to d_I \bar d_J) ~ {\rm when}~ 
\left( \lambda^\prime_{i3I} \times \lambda^\prime_{i3J} \right)^2 \neq 0~;~
i=1,2,3 \label{br2} \\
BR3^{IJ} & = & 
BR(Z \to d_I \bar d_J) ~ {\rm when}~  
\left( \lambda^\prime_{iIj} \times \lambda^\prime_{iJj} \right)^2 \neq 0 
~;~i,j=1,2,3 \label{br3}~,
\end{eqnarray} 
\end{widetext}

\noindent such that $BR(Z \to b \bar s)=BR1^{32},~BR2^{32}$ 
or $BR3^{32}$ depending on which of the three $\lambda^\prime \lambda^\prime$ 
product combinations is non-zero.

\begin{table*}[htb]
\begin{center}
%\begin{tabular}{||c|c|c|c||}
\begin{tabular}{c|c|c|c}
%\hline
%%%%%%%%%%%%%%%%%%%%%%%%%%%%%%%%%%%%%%%%%%%%%%%%%%%%
 & $\frac{BR1^{32}}{\left( \lambda^\prime_{ij3} \times \lambda^\prime_{ij2} \right)^2} ~~, j \ne 3$  & 
$\frac{BR2^{32}}{\left( \lambda^\prime_{i33} \times \lambda^\prime_{i32} \right)^2}$ &
$\frac{BR3^{32}}{\left( \lambda^\prime_{i3j} \times \lambda^\prime_{i2j} \right)^2} $ 
\\ 
~&~&~&\\
\hline
~&~&~&\\
$m_{\tilde q}=500$ GeV,  
$m_{\tilde\ell}=200$ GeV &  
$4.2 \times 10^{-7}$ & 
$2.4 \times 10^{-6}$ & 
$3.4 \times 10^{-6}$ 
\\ 
~&~&~&\\
%\hline
$m_{\tilde q}=1000$ GeV,  
$m_{\tilde\ell}=500$ GeV &  
$3.9 \times 10^{-7}$ & 
$6.4 \times 10^{-8}$ & 
$3.0 \times 10^{-6}$ 
\\ 
~&~&~&\\
%\hline
\end{tabular}
\caption{Results for the three types of branching ratios 
 $BR1^{32},~BR2^{32},~BR3^{32}$ as defined in (\ref{br1})-(\ref{br3}), 
each scaled by its appropriate 
 $\lambda^\prime \lambda^\prime$ coupling product. Results are given 
for two sets of squark and slepton
masses as indicated.}
\label{tabA}
\end{center}
\end{table*}

In Table \ref{tabA} we give a sample of our numerical results 
for the three BR's in (\ref{br1})-(\ref{br3}) scaled by the
square of the appropriate $\lambda^\prime \lambda^\prime$ product. 
The results presented 
in Table \ref{tabA} correspond to
the case of a single non-zero 
$\lambda^\prime \lambda^\prime$ product (one index combination) 
contributing to each of the BR's  
$BR1^{32},~BR2^{32}$ and $BR3^{32}$. 
In addition, 
the masses of the squark and slepton being exchange in the loop
(for a given index combination 
of the corresponding $\lambda^\prime \lambda^\prime$ product)  
are set to either 
$m_{\tilde q}=500$ GeV with  
$m_{\tilde\ell}=200$ GeV or $m_{\tilde q}=1000$ GeV with  
$m_{\tilde\ell}=500$ GeV.

The existing limits on the $\lambda^\prime$ coupling products
in (\ref{br1})-(\ref{br3})  
seem to indicate that the typical allowed values of  
$\lambda^\prime \times \lambda^\prime$ 
for any of the index combinations in (\ref{br1})-(\ref{br3}) is
at the level of $\sim {\rm few} \times 10^{-2}$ 
\cite{limlamlam}. It should be noted, however, that the 
limits 
reported in \cite{limlamlam} assume 
100 GeV scalar masses. 
These limits scale with the scalar 
masses (typically as $[m_{\tilde s}/100~{\rm GeV}]^2$, where $m_{\tilde s}$ 
is the scalar mass) 
and are, therefore,
relaxed as the scalars become heavier.$^{[8]}$\footnotetext[8]{Note 
that $b \to s \gamma$, which is proportional to 
$\lambda^\prime \times \lambda^\prime$ products with the same index 
combinations as in $Z \to b \bar s$, allows some of the above 
$\lambda^\prime \times \lambda^\prime$ coupling products to be
at the $10^{-1}$ level \cite{bsglamlamlim}.}
 
Using $\lambda^\prime \times \lambda^\prime \sim {\cal O}(10^{-2})$ 
in conjunction with the results presented in 
Table \ref{tabA}, we see that the expected branching ratio
for $Z \to b \bar s$ in the type A RPV scenario investigated 
in this section lies in the range 
$BR(Z \to d_I \bar d_J) \sim 10^{-11} - 10^{-10}$. 

This type A 
RPV one-loop effect in $BR(Z \to d_I \bar d_J)$ was also 
investigated in \cite{shemtob}.  
Although \cite{shemtob} evaluated some distinct limiting cases 
of the type A RPV contributions, our results agree 
with the highlights of their analysis, i.e., that 
the typical $BR(Z \to d_I \bar d_J)$ is expected to be at the 
level of $10^{-11} - 10^{-10}$ if 
$\lambda^\prime \times \lambda^\prime \sim {\cal O}(10^{-2})$.

Thus, the type A RPV scenario is expected to yield 
a BR smaller even from the SM one.   
We, therefore, proceed below to the second RPV type B scenario which 
seems to give a much larger $BR(Z \to b \bar s)$ within the 
experimentally allowed
range of values for its relevant RPV parameter space.

\subsection{Type B RPV effect}

The type B RPV effect arises 
when a Higgs particle that is being exchanged in the loops  
mixes with a slepton through 
the RPVB operator in (\ref{bterm}) and then couples
to the external down quark via a $\lambda^\prime$ type coupling.

For simplicity we will assume that $b_i \neq
0$ only for $i=3$ in (\ref{bterm}), thus, considering only the 
mixing between the third generation sleptons 
($\tilde L_3$) 
and the Higgs scalar fields ($H_d$ and $H_u$).$^{[9]}$
It should be noted that $b_3 \neq 0$ leads in general to 
a non-vanishing tau-sneutrino VEV, $v_3$.
However, since lepton number is not a conserved quantum number
in this scenario, the $\hat H_d$ and 
$\hat L_3$ superfields lose their identity and can be rotated to a
particular basis ($\hat H_d^\prime,\hat L_3^\prime$) in which
either $\mu_3$ or $v_3$ are set to zero
\cite{GH,davidson,basis,davidsonnew}. In what follows, we
find it convenient to choose the ``no VEV'' basis, $v_3=0$, which
simplifies our analysis below.

Let us define the SU(2) components of the up-type Higgs, down-type 
Higgs and $\tilde L_3$ scalar doublet fields (setting $v_3=0$):

\begin{eqnarray}
H_{u} &\equiv& \pmatrix{h_u^+ \cr (\xi_{u}^0 + v_{u} + 
i \phi_{u}^0)/\sqrt{2}} ~,\nonumber \\
H_{d} &\equiv& \pmatrix{(\xi_{d}^0 + v_{d} + i \phi_{d}^0)/\sqrt{2} \cr
h_d^-}
\label{components}~, \\
\tilde L_3 &\equiv& \pmatrix{(\tilde\nu_{+}^0 + i
\tilde\nu_{-}^0)/\sqrt{2} \cr \tilde e_{3}^-} \nonumber ~,
\end{eqnarray}

\noindent where $\tilde\nu_{+}^0$, $\tilde\nu_{-}^0$ 
and $\tilde e_{3}^-$
are 
the SU(2) CP-even, CP-odd $\tau$-sneutrino and  
left handed stau fields, respectively.

When $b_3 \neq 0$ the 3rd generation 
slepton SU(2) fields in (\ref{components}) 
mix with the Higgs fields.
In particular, in the basis $\Phi_C^0 = (h_u^+,h_d^+,\tilde e_{3}^+)$,
the squared mass matrix in the charged scalar sector becomes:$^{[10]}$

\begin{widetext}
\begin{eqnarray}
M^2_C = \pmatrix{c_\beta^2 \left[ (m_A^0)^2 + m_W^2 \right] &
s_\beta c_\beta  \left[ (m_A^0)^2 + m_W^2 \right]  &  b_3 \cr 
 s_\beta c_\beta  \left[ (m_A^0)^2 + m_W^2 \right] & 
s_\beta^2  \left[ (m_A^0)^2 + m_W^2 \right] & b_3 t_\beta \cr 
b_3 & b_3 t_\beta & (m_{s \nu}^0)^2 - m_W^2 \cos 2 \beta } \label{mc2}~,
\end{eqnarray}
\end{widetext}

\noindent where $m_A^0$ and $m_{s\nu}^0$ are 
the pseudo-scalar Higgs mass and the tau-sneutrino mass, respectively,
in the RPC limit $b_3  \to 0$. 

Similarly, in the basis $\Phi_E^0 = (\xi_{d}^0,\xi_{u}^0,\tilde\nu_{+}^0)$, 
the CP-even neutral scalar squared mass matrix 
becomes (at tree-level):$^{[11]}$

\begin{widetext}
\begin{eqnarray}
M^2_E = \pmatrix{(m_A^0)^2 s_\beta^2 + m_Z^2 c_\beta^2 & 
-\left[ (m_A^0)^2 + m_Z^2 \right] s_\beta c_\beta & b_3 t_\beta \cr 
-\left[ (m_A^0)^2 + m_Z^2 \right]
s_\beta c_\beta & (m_A^0)^2 c_\beta^2 + m_Z^2
s_\beta^2 & -b_3 \cr b_3 t_\beta & -b_3 & (m_{s
\nu}^0)^2} \label{me2}~.
\end{eqnarray} 
\end{widetext}

\noindent Finally, in the CP-odd neutral scalar sector and in the basis
$\Phi_O^0 = (\phi_{d}^0,\phi_{u}^0,\tilde\nu_{-}^0)$ one obtains:

\begin{eqnarray}
M^2_O = \pmatrix{(m_A^0)^2 c_\beta^2 & (m_A^0)^2 c_\beta s_\beta &
b_3 t_\beta \cr (m_A^0)^2 c_\beta s_\beta & (m_A^0)^2 s_\beta^2 &
 b_3 \cr
 b_3 t_\beta & b_3 & (m_{s \nu}^0)^2 } \label{mo2}~.
\end{eqnarray}

\n The new charged scalar and CP-even and CP-odd neutral scalar 
mass-eigenstates (i.e., the physical states)  are then derived by
diagonalizing $M_C^2,~M_{E}^2$ and $M_O^2$, respectively.
Let us denote the physical states by:
\footnotetext[9]{The 
consequences of $b_1 \neq 0$ and/or $b_2 \neq 0$ is to
introduce additional mixings among sleptons of different
generations and mixings between the selectron and/or smuon scalar 
doublets with
the Higgs fields. These extra mixing effects are not crucial for 
the main outcome of this section.}
\footnotetext[10]{We 
neglect the mixing between the  right-handed SU(2) stau singlet and the 
charged Higgs fields
which is proportional to the tau mass.}
\footnotetext[11]{The one loop corrections 
to the $2 \times 2$ Higgs block in $M_E^2$ can cause a significant 
deviation to the tree-level mass of the light Higgs. This effect 
will be discussed below.}

\begin{equation}
\Phi_C = \pmatrix{H^+ \cr G^+ \cr \tilde\tau^+}~,~
\Phi_E =\pmatrix{H \cr h \cr \tilde\nu_+^\tau} ~,~
\Phi_O = \pmatrix{A \cr G \cr \tilde\nu_-^\tau} \label{newstates}~,
\end{equation}

\noindent such that, for a small RPVB in the scalar potential, 
the new physical states in (\ref{newstates}) 
are the states dominated by what would be 
the corresponding physical states in the RPC limit, $b_3=0$, for 
which the Higgs sector decouples from the slepton sector in 
(\ref{mc2}), (\ref{me2}) and (\ref{mo2}).
In particular, if $b_3\to 0$, then $H,~h,~A$ and $H^+$ 
become the usual RPC MSSM's CP-even heavy Higgs, 
CP-even light Higgs, CP-odd pseudo-scalar Higgs and 
charged Higgs states, respectively.    
Similarly, in this limit $\tilde\nu_+^\tau$ and $\tilde\nu_-^\tau$ become
the two mass-degenerate CP-even and CP-odd sneutrino states with
a common mass 
$m_{\tilde\nu_+^\tau}=m_{\tilde\nu_-^\tau} \equiv m_{s\nu}^0$, while 
$\tilde\tau^+$ is the usual pure left handed stau field with a mass 
$m_{\tilde\tau^+}=\sqrt{(m_{s \nu}^0)^2 - m_W^2 \cos 2 \beta}$. 
Note also that $G$ and $G^+$ are the neutral and charged  
Goldstone bosons that are absorbed by the $Z$ and $W$-bosons 
and are, therefore, the states with a zero eigenvalue in $M_O^2$ and 
$M_C^2$, respectively.

The physical states $\Phi_{C,}$, $\Phi_{E}$ and 
$\Phi_{O}$ are related to the weak
states $\Phi_{C}^0$, $\Phi_{E}^0$ and $\Phi_{O}^0$ via:

\begin{eqnarray}
\Phi^0_{C,i} &=& R_{C,ik}\Phi_{C,k}~,\nonumber\\
\Phi^0_{E,i} &=& R_{E,ik} \Phi_{E,k} \label{typebrot}~,\\
\Phi^0_{O,i} &=& R_{O,ik}\Phi_{O,k} ~,\nonumber
\end{eqnarray}

\n where $R_{C}$, $R_{E}$ and $R_{O}$ 
are the rotation matrices that
diagonalize $M_{C}^2$, $M_{E}^2$ and $M_{O}^2$, respectively.

Notice that the mass matrices 
$M_C^2,~M^2_E$ and $M_O^2$ depend only on four SUSY parameters: 
$m_A^0,~m_{s\nu}^0,b_3$ and $\tan\beta$. These parameters, therefore,  
completely fix the rotation matrices $R_C,~R_O$ and $R_E$ from which 
the CP-even and CP-odd neutral scalar spectrum as well as the charged 
scalar spectrum is completely determined.

Clearly then, the type B RPV contributions involve 
the 3rd generation sleptons that can mix with the Higgs fields 
through a $b_3$ bilinear RPV coupling which enters the 
slepton-Higgs mixed mass matrices in (\ref{mc2})-(\ref{mo2}).
Here also, we can further sub-divide the type B RPV effects
according to the type of scalar ($S$) and type of fermion ($f$)
that are being exchanged in the loops:
%\begin{eqnarray}
%{\rm type~B1:} && S_\alpha=\Phi_{C,\alpha},~f_i=u_i ~;~~~~~~~~~~~~~~
%\alpha=1,3~{\rm and}~ i=1,2,3 ~, \nonumber \\
%{\rm type~B2:} && S_\alpha=\Phi_{E,\alpha}~{\rm and}~\Phi_{O,\beta},~
%f_i=d_i ~;~
%\alpha=1,2,3~,\beta=1,3~{\rm and}~i=1,2,3 \label{typeb}~.
%\end{eqnarray}
\begin{equation}
{\rm type~B1:}~~ S_\alpha=\Phi_{C,\alpha}~;~~f_i=u_i \label{typeb1}~,
\end{equation}

\n with $\alpha=1,3$, $i=1,2,3$ and 
\begin{equation}
{\rm type~B2:} ~~ S_\alpha=\Phi_{E,\alpha}~{\rm and}~\Phi_{O,\beta}~;~~
f_i=d_i \label{typeb2}~,
\end{equation}

\n with $\alpha=1,2,3~,\beta=1,3$, $i=1,2,3$.

Note that we have omitted virtual exchanges 
of the two Goldstone bosons $G$ and $G^+$ since 
the one-loop amplitudes are being calculated in the unitary gauge. 

The two RPV effects, of types B1 and B2 above, 
are driven by 
the Higgs-slepton scalar admixtures $\Phi_{C}$, $\Phi_{E}$, 
$\Phi_{O}$ which 
couple to quarks through a combination of $\lambda^\prime$ and Higgs
Yukawa coupling. Hence, the Higgs-like components in $\Phi_{C}$, 
$\Phi_{E}$ and $\Phi_{O}$
will couple through the Higgs Yukawa terms, while the slepton-like component 
interact with the quarks via the $\lambda^\prime$-type RPV couplings 
in (\ref{rpv}).

For the type B1 RPV contribution in (\ref{typeb1}) 
the form factors defined 
in (\ref{muk}) are calculated following the prescription described 
in section II. The generic couplings defined in (\ref{Vff}), 
(\ref{VSS}) and (\ref{Sdf}) 
are summarized in Table \ref{tab4} for the type B1 RPV exchanges.
In particular, the $Sdf$ couplings (for $S=\Phi_{C}$ and $f=u$)
are a combination of  
the Yukawa-like trilinear RPV interactions in (\ref{rpv}) 
(those with the third generation slepton indices) 
and the charged Higgs Yukawa couplings which are the same as in 
the 2HDM of type II (given in section III).  

The $ZSS$  couplings (for $S=\Phi_C$) in Table \ref{tab4} are 
derived from ${\cal L}(V_\mu \tilde L_3 \tilde L_3)$ in (\ref{velel})
and from the following 
${\cal L}(V_\mu H_d H_d)$ and ${\cal L}(V_\mu H_u H_u)$ 
pieces \cite{rosiek}: 

\begin{equation}
{\cal L}(V_\mu H_d H_d) = -i \frac{1}{2}
 H_d^{\star} \left[  \frac{e}{s_W}
 \tau^3 A_\mu^3 -  \frac{e}{c_W} B_\mu \right] 
 \stackrel{\leftrightarrow}{\partial^\mu_{\tilde H}} 
H_d \label{Vhdhd}~,
\end{equation}
\begin{equation}
{\cal L}(V_\mu H_u H_u) = -i \frac{1}{2}
 H_u^{\star} \left[  \frac{e}{s_W}
 \tau^3 A_\mu^3 + \frac{e}{c_W} B_\mu \right] 
 \stackrel{\leftrightarrow}{\partial^\mu_{\tilde H}} 
H_u \label{Vhuhu}~,
\end{equation}

\noindent where the SU(2) scalar doublet fields $\tilde L_3,~H_d$ and $H_u$ 
are defined in (\ref{components}).

\begin{table}[htb]
\begin{center}
%\begin{tabular}{||c|c||}
\begin{tabular}{c|c}
%\hline
%%%%%%%%%%%%%%%%%%%%%%%%%%%%%%%%%%%%%%%%%%%%%%%%%%%%
 & type B1
\\ 
~ \\
\hline
~ \\
scalar ($S_{\alpha}$) & 
$\Phi_{C,\alpha},~~\alpha=1,3$ 
\\ 
~ \\
%\hline
fermion ($f_i$) & 
$u_i,~~i=1,2,3$
\\ 
~ \\
%\hline
$a_{L(Zf)}^{ij}$ & 
$a_{L(Zu)}$ 
\\
~ \\
%\hline
$a_{R(Zf)}^{ij}$ & 
$a_{R(Zu)}$ 
\\ 
~ \\
%\hline
$b_{L(\alpha)}^{ij}$ & 
$\frac{e}{\sqrt{2}s_W} \frac{m_{u_i}}{m_W s_\beta} R_C^{1 \alpha}V_{ij}$
\\ 
~ \\
%\hline
$b_{R(\alpha)}^{ij}$ & 
$\frac{e}{\sqrt{2}s_W} \frac{m_{d_j}}{m_W c_\beta} R_C^{2 \alpha}V_{ij} 
-\lambda_{3ij}^{\prime \star} R_C^{3 \alpha}$
\\ 
~ \\
%\hline
$g_Z^{\alpha \beta}$ &
$- e \cot 2 \theta_W  \delta_{\alpha \beta}$
\\ 
~ \\
%\hline
\end{tabular}
\caption{The couplings required for the calculation 
of $\Gamma(Z\to d_I \bar d_J)$ in the type B1 RPV scenario. 
The couplings follow the notation in (\ref{Vff}), (\ref{VSS}) and 
(\ref{Sdf}). The couplings $a_{L,R(Zu)}$ are given in 
(\ref{Vdd})-(\ref{Vddlast}).  
\label{tab4}}
\end{center}
\end{table}

\begin{figure*}[htb]
\psfull
 \begin{center}
\vspace{0.5cm}
  \leavevmode
  \epsfig{file=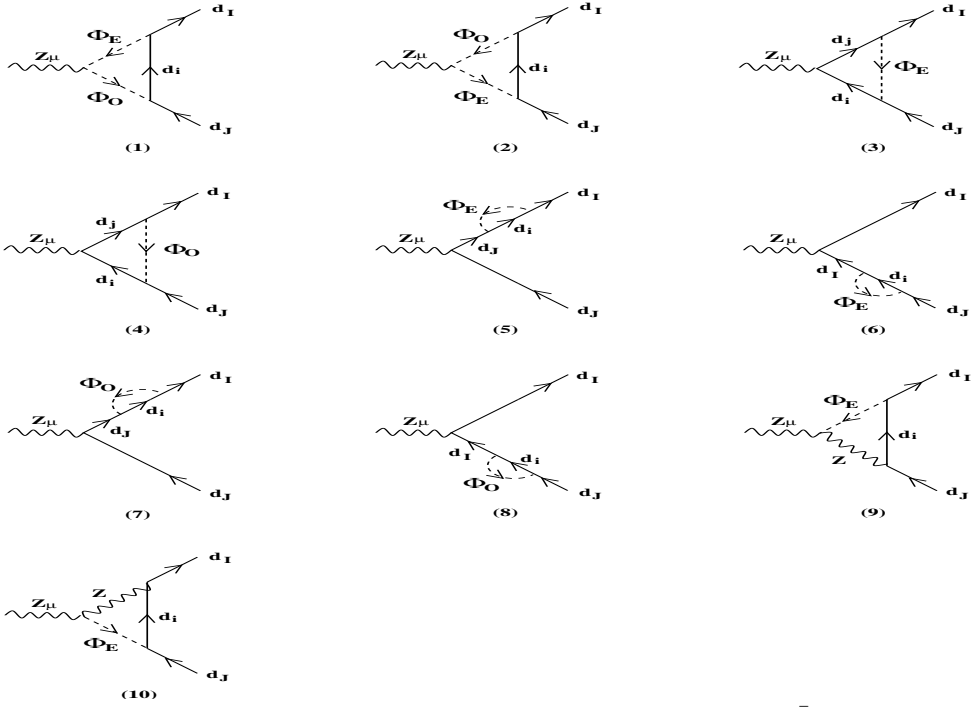,height=8cm,width=14cm,bbllx=0cm,bblly=2cm,bburx=20cm,bbury=25cm,angle=0}
 \end{center}
\caption{One-loop diagrams that contribute to the 
flavor changing decays $Z \to d_I \bar d_J$ in the type B2 
RPV scenario.}
\label{fig2}
\end{figure*}

For the type B2 RPV case  (see (\ref{typeb2})) there are 10 one-loop diagrams 
that can potentially contribute to 
the decay $Z\to d_I \bar d_J$. 
These diagrams are depicted in Fig.~\ref{fig2}. 
The first eight diagrams in Fig.~\ref{fig2} have the same topology as 
the generic diagrams of Fig.~\ref{fig1}, while diagrams 9 and 10 
involve virtual exchanges of a Z-boson through 
the $ZZ\Phi_E$ interaction. 

Using our generic notation for the one-loop 
amplitude in (\ref{muk}), we calculate 
the form factors $A_{L,k}^{IJ},~A_{R,k}^{IJ}$ 
and $B_{L,k}^{IJ},~B_{R,k}^{IJ}$ with $k=1-10$, 
which emerge from diagrams 1-10 in Fig.~\ref{fig2} 
(taking $m_{d_i}=0$, for $d,~s$ and $b$-quarks):

\begin{eqnarray}
A^{IJ}_{L,1} &=& 
-2 \sum_{\alpha,\beta,i} g_{Z}^{\alpha \beta} b_{L(\alpha)}^{(E)iI} 
 b_{L(\beta)}^{(O)iJ} C_{24}^{1} \label{form1}~,  \\
A^{IJ}_{L,2} &=& 
2 \sum_{\alpha,\beta,i} g_{Z}^{\beta \alpha} b_{L(\alpha)}^{(O)iI} 
 b_{L(\beta)}^{(E)iJ} C_{24}^{2} ~,  \\
A^{IJ}_{L,3} &=& a_{R(Zd)} \sum_{\alpha,i,j} b_{L(\alpha)}^{(E)iI}  
b_{L(\alpha)}^{(E)jJ}  
\left[ 2 C_{24}^{3} \right. \nonumber \\
&& \left. - m_Z^2  \left( C_{23}^{3} - C_{22}^{3} 
\right)  \right] ~, \\
A^{IJ}_{L,4} &=& a_{R(Zd)} \sum_{\alpha,i,j} b_{L(\alpha)}^{(O)iI}  
b_{L(\alpha)}^{(O)jJ} 
\left[ 2 C_{24}^{4} \right. \nonumber \\ 
&& \left. - m_Z^2  \left( C_{23}^{4} - C_{22}^{4} 
\right)  \right] ~, \\
A^{IJ}_{L,56} &=& a_{L(Zd)} \sum_{\alpha,i}
b_{L(\alpha)}^{(E)iI}  b_{L(\alpha)}^{(E)iJ} B_1^{5} ~,\\
A^{IJ}_{L,78} &=& a_{L(Zd)} \sum_{\alpha,i}
b_{L(\alpha)}^{(O)iI}  b_{L(\alpha)}^{(O)iJ} B_1^{7} ~,\\
A^{IJ}_{L,9} &=& A^{IJ}_{L,10} =0 ~,
\end{eqnarray}

\noindent where we have combined the contribution of the 
self energy diagrams 5+6 and 7+8: 
${\cal M}_5+{\cal M}_6 \equiv {\cal M}_{56}$ and 
${\cal M}_7+{\cal M}_8 \equiv {\cal M}_{78}$, which 
leads accordingly to 
$A^{IJ}_{L,5} + A^{IJ}_{L,6} \equiv A^{IJ}_{L,56}$ and 
$A^{IJ}_{L,7} + A^{IJ}_{L,8} \equiv A^{IJ}_{L,78}$. Also,

\begin{eqnarray}
B^{IJ}_{L,k} &=& 0 ~ {\rm for}~ k=1-8~\\
B^{IJ}_{L,9} &=& a_{L(Zd)} \sum_{\alpha} g_{ZZ\Phi_E}^{\alpha}
b_{L(\alpha)}^{(E)IJ} \left[2 \left( C_{12}^{9} - C_{11}^{9} \right) 
\right. \nonumber \\
&& \left.
+\frac{1}{m_Z^2} \left( \tilde C_{0}^{9} + \tilde C_{11}^{9} \right)
\right] ~,\\
B^{IJ}_{L,10} &=& a_{R(Zd)} \sum_{\alpha} g_{ZZ\Phi_E}^{\alpha}
b_{L(\alpha)}^{(E)IJ} \left[2 \left( C_{12}^{10} - C_{11}^{10} \right) 
\right. \nonumber \\
&& \left. + \frac{1}{m_Z^2} \left( \tilde C_{0}^{10} + 
\tilde C_{11}^{10} \right)
\right] \label{form2}~.
\end{eqnarray}

\noindent Here also the right-handed form factors, 
$A^{IJ}_{R,k}$ and $B^{IJ}_{R,k}$, are obtained from the corresponding
left handed ones  
by interchanging $L \to R$ and $R \to L$ in all the 
couplings in (\ref{form1})-(\ref{form2}).

The two-point and three-point loop form factors $B_1^k$ with 
$k=5,7$, $C_x^k$ with $x \in 11,12,21,22,23,24$ and 
$k=1,2,3,4,9,10$ and 
$\tilde C_x^k$ with $x \in 0,11,12$ and 
$k=9,10$
which appear in (\ref{form1})-(\ref{form2}) are given 
by:

\begin{eqnarray}
B_1^{5} &=& B_1 \left(m_{d_i}^2,m_{\Phi_{E,\alpha}}^2,m_{d_I}^2 \right) ~,\\
B_1^{7} &=& B_1 \left(m_{d_i}^2,m_{\Phi_{O,\alpha}}^2,m_{d_I}^2 \right) ~,
\end{eqnarray}

\noindent and 
  
\begin{equation}
C_x^{1} = 
C_x \left(m_{d_i}^2,m_{\Phi_{E,\alpha}}^2,m_{\Phi_{O,\beta}}^2,
m_{d_I}^2,q^2,m_{d_J}^2 \right) ~,
\end{equation}
\begin{equation}
C_x^{2} =
C_x \left(m_{d_i}^2,m_{\Phi_{O,\alpha}}^2,m_{\Phi_{E,\beta}}^2,
m_{d_I}^2,q^2,m_{d_J}^2 \right) ~,
\end{equation}
\begin{equation}
C_x^{3} =
C_x \left(m_{\Phi_{E,\alpha}}^2,m_{d_i}^2,m_{d_j}^2,m_{d_J}^2,q^2,m_{d_I}^2 
\right) ~,
\end{equation}
\begin{equation}
C_x^{4} = 
C_x \left(m_{\Phi_{O,\alpha}}^2,m_{d_i}^2,m_{d_j}^2,m_{d_J}^2,q^2,m_{d_I}^2 
\right) ~,
\end{equation}
\begin{equation}
C_x^{9};~\tilde C_x^9 = 
C_x;~\tilde C_x \left(m_{d_J}^2,m_Z^2,m_{\Phi_{E,\alpha}}^2,
m_{d_J}^2,q^2,m_{d_I}^2 \right) ~,
\end{equation}
\begin{equation}
C_x^{10};~\tilde C_x^{10} =
C_x;~\tilde C_x \left(m_{d_I}^2,m_Z^2,m_{\Phi_{E,\alpha}}^2,
m_{d_I}^2,q^2,m_{d_J}^2 \right) ~,
\end{equation}

\noindent where $B_1(m_1^2,m_2^2,p^2)$, 
$C_x(m_1^2,m_2^2,m_3^2,p_1^2,p_2^2,p_3^2)$ and 
$\tilde C_x(m_1^2,m_2^2,m_3^2,p_1^2,p_2^2,p_3^2)$ are defined in the 
appendix. 

The couplings $a_{L(Zd)}$, $a_{R(Zd)}$, 
$b_{L(\alpha)}^{(E)ij}$,
$b_{R(\alpha)}^{(E)ij}$,
$b_{L(\alpha)}^{(O)ij}$,
$b_{R(\alpha)}^{(O)ij}$,
$g_{Z}^{\alpha \beta}$ and  
$g_{ZZ\Phi_E}^{\alpha}$ needed for evaluating 
the form factors above are given in Table \ref{tab5}.
In particular, $a_{L,R(Zd)}$ are the SM left and right-handed couplings 
of the Z-boson to a pair of down quarks as given in 
(\ref{Vdd})-(\ref{Vddlast}). 
The rest are
obtained from the relevant 
interaction Lagrangian terms by rotating the SU(2) weak states 
$\Phi^0_{C,E,O}$ to the physical states $\Phi_{C,E,O}$ according 
to (\ref{typebrot}). In particular,
the $\Phi_E \bar d_i d_j$ couplings 
$b_{L,R(\alpha)}^{(E)ij}$ and $\Phi_O \bar d_i d_j$ couplings 
$b_{L,R(\alpha)}^{(O)ij}$ follow the notation of the generic 
$S d f$ vertex in (\ref{Sdf}); for $S=\Phi_E$ or $\Phi_O$ 
and $f=d$:

\begin{eqnarray}
\Lambda(\Phi_{E,\alpha} \bar d_i d_j) &=& i \left(b_{L(\alpha)}^{(E)ij} L+
b_{R(\alpha)}^{(E)ij} R \right) \label{bLRE}~, \\
\Lambda(\Phi_{O \alpha} \bar d_i d_j) &=& i \left(b_{L(\alpha)}^{(O)ij} L+
b_{R(\alpha)}^{(O)ij} R \right) \label{bLRO}~.
\end{eqnarray}

\noindent These couplings emanate from both the 
Yukawa-like trilinear RPV interactions in (\ref{rpv})  
and the neutral Higgs Yukawa vertices
of a 2HDM of type II as given in (\ref{yukawa}). 

The coupling $g_{Z}^{\alpha \beta}$
of a $Z$-boson to a $\Phi_{E,\alpha} \Phi_{O,\beta}$ pair 
follow our generic definition of the $VSS$ vertex 
in (\ref{VSS}). It is derived 
from the Lagrangian terms in (\ref{velel}), (\ref{Vhdhd}) and 
(\ref{Vhuhu}). 
 
The coupling $g_{ZZ\Phi_E}^{\alpha}$ of $\Phi_{E,\alpha}$ to a 
pair of $Z$-bosons is defined as:

\begin{eqnarray}
\Lambda(Z_\mu Z_\nu \Phi_{E,\alpha})= i g_{ZZ\Phi_E}^{\alpha} g_{\mu \nu}
\label{zzphi}~,
\end{eqnarray}

\n and is obtained from 
the following $ZZ\xi_d^0$ and $ZZ \xi_u^0$ interaction terms (recall that 
$\xi_{d,u}^0$ are the CP-even SU(2) components of $H_{d,u}$ as defined in 
(\ref{components})) \cite{rosiek}:

\begin{eqnarray}
{\cal L}(ZZ \xi_{d,u}^0)=\frac{e^2}{(2 s_W c_W)^2} Z_\mu Z^\mu 
\left( v_d \xi_d^0 + v_u \xi_u^0 \right) \label{ZZE}~,
\end{eqnarray}

\noindent where $v_d$ and $v_u$ are the VEV's of the down-type and up-type
Higgs doublets, respectively.  

\begin{table}[htb]
\begin{center}
%\begin{tabular}{||c|c||}
\begin{tabular}{c|c}
%\hline
%%%%%%%%%%%%%%%%%%%%%%%%%%%%%%%%%%%%%%%%%%%%%%%%%%%%
 & type B2
\\ 
~ \\
\hline
~ \\
scalar ($S_{\alpha}$) & 
$\Phi_{E,\alpha},~~\alpha=1,2,3$ and $\Phi_{O,\alpha},~~\alpha=1,3$ 
\\ 
~ \\
%\hline
fermion ($f_i$) & 
$d_i,~~i=1,2,3$
\\ 
~ \\
%\hline
$a_{L(Zf)}^{ij}$ & 
$a_{L(Zd)}$  
\\
~ \\
%\hline
$a_{R(Zf)}^{ij}$ & 
$a_{R(Zd)}$
\\ 
~ \\
%\hline
$b_{L(\alpha)}^{(E)ij}$ & 
$-\frac{e}{2 s_W} \frac{m_{d_i}}{m_W c_\beta} R_E^{1 \alpha} \delta_{ij}
+\frac{1}{\sqrt{2}} \lambda_{3ji}^{\prime} R_E^{3 \alpha}  $ 
\\ 
~ \\
%\hline
$b_{R(\alpha)}^{(E)ij}$ & 
$-\frac{e}{2 s_W} \frac{m_{d_i}}{m_W c_\beta} R_E^{1 \alpha} \delta_{ij}
+\frac{1}{\sqrt{2}} \lambda_{3ij}^{\prime \star} R_E^{3 \alpha} $
\\ 
~ \\
%\hline
$b_{L(\alpha)}^{(O)ij}$ & 
$ -i \frac{e}{2 s_W} \frac{m_{d_i}}{m_W c_\beta} R_O^{1 \alpha} \delta_{ij}
+\frac{i}{\sqrt{2}} \lambda_{3ji}^{\prime} R_O^{3 \alpha}  $ 
\\ 
~ \\
%\hline
$b_{R(\alpha)}^{(O)ij}$ & 
$i \frac{e}{2 s_W} \frac{m_{d_i}}{m_W c_\beta} R_O^{1 \alpha} \delta_{ij}
+\frac{i}{\sqrt{2}} \lambda_{3ij}^{\prime \star} R_O^{3 \alpha} $ 
\\ 
~ \\
%\hline
$g_Z^{\alpha \beta}$ &
$i \frac{e}{\sin 2 \theta_W} \left( R_E^{1 \alpha} R_O^{1 \beta}-
R_E^{2 \alpha} R_O^{2 \beta}+
R_E^{3 \alpha} R_O^{3 \beta} \right)$ 
\\ 
~ \\
%\hline
$g_{ZZ \Phi_E}^{\alpha}$ &
$\frac{e}{s_W c_W} m_Z \left( c_\beta R_E^{1 \alpha}
+ s_\beta R_E^{2 \alpha} \right)$ 
\\ 
~ \\
%\hline
\end{tabular}
\caption{The couplings required for the calculation 
of $\Gamma(Z\to d_I \bar d_J)$ in the type B2 RPV scenario. 
The couplings follow from the 
Feynman rules in (\ref{Vff}), (\ref{VSS}), (\ref{bLRE}), (\ref{bLRO}) 
and (\ref{zzphi}). 
The couplings $a_{L,R(Zd)}$ are given in 
(\ref{Vdd})-(\ref{Vddlast}).  
\label{tab5}}
\end{center}
\end{table}

Before presenting our numerical results for the type B RPV contribution
let us discuss some of its salient features 
and outline the main assumptions and notations
regarding the relevant parameter space involved:

\begin{description}

\item{I.} The pseudo-scalar ``bare'' 
mass (i.e., its mass in the RPC limit of $b_3 \to 0$) can
be approximated from 
the tree-level relation which, for $t_\beta^2 \gg 1$, gives 
$(m_A^0)^2 \sim b_0 t_\beta$, where $b_0$ is the usual 
RPC soft-breaking bilinear Higgs term in the
scalar potential, i.e., $V_{RPC} \supset b_0 H_d H_u$.

Thus, without loss of generality, 
we trade the bilinear coupling $b_3$ with a dimensionless 
RPV parameter, $\varepsilon$, as follows:

\begin{eqnarray}
b_3 \equiv \varepsilon (m_A^0)^2 \cot\beta \label{b3toeps}~,
\end{eqnarray}

\n such that $\varepsilon \sim \frac{b_3}{b_0}$ parametrizes the relative 
amount of RPV in the scalar potential. 
In particular, $\varepsilon \ll 1$ for small bilinear RPV  
and $\varepsilon \sim 1$ if RPV/RPC $\sim 1$ in the SUSY scalar sector.

The existing experimental limits on $\varepsilon$ come from:

\begin{figure*}[htb]
\psfull
 \begin{center}
  \leavevmode
  \epsfig{file=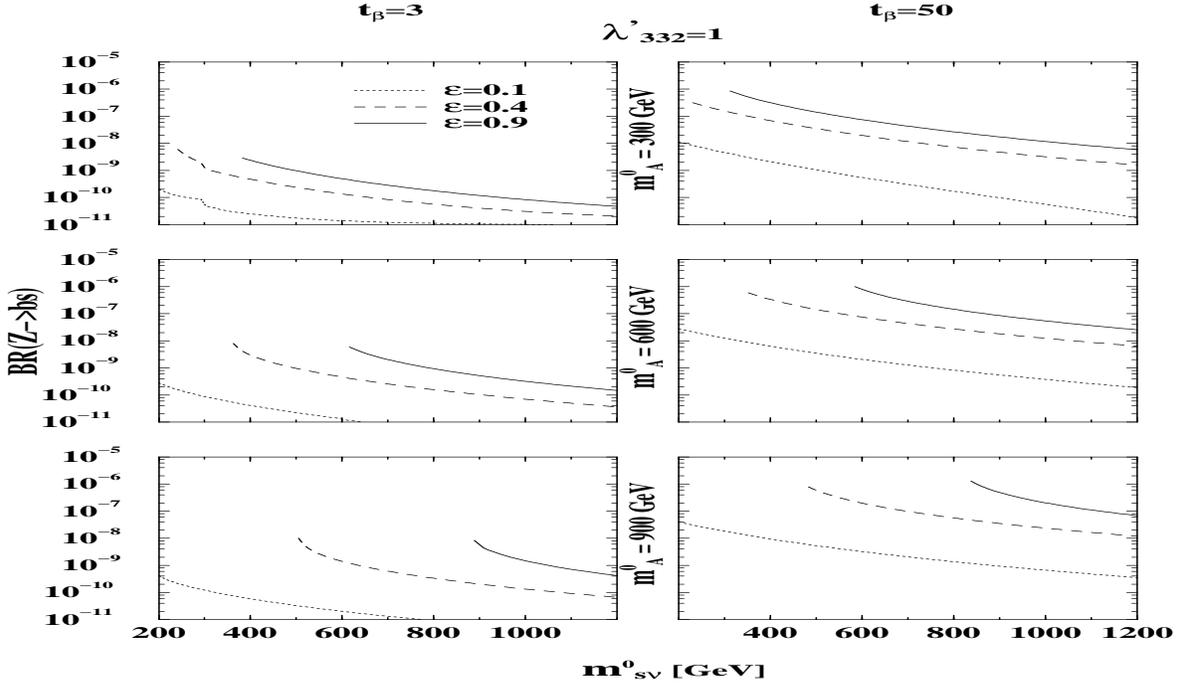,height=8cm,width=16cm,bbllx=0cm,bblly=2cm,bburx=20cm,bbury=25cm,angle=0}
 \end{center}
\caption{$BR(Z\to b \bar s + \bar b s)$ as a function 
of the ``bare'' sneutrino mass parameter $m_{s \nu}^0$ (see text), 
for some combinations of values of $m_A^0$ and $\varepsilon$ (as indicated 
in the figure) and for $t_\beta=3$ (left plots) and $t_\beta=50$ 
(right plots). $\lambda^\prime_{332}=1$ is used and $\epsilon$ is defined 
in (\ref{b3toeps}).}
\label{figtypeB1}
\end{figure*}

\begin{figure*}[htb]
\psfull
 \begin{center}
\vspace{1.cm}
  \leavevmode
  \epsfig{file=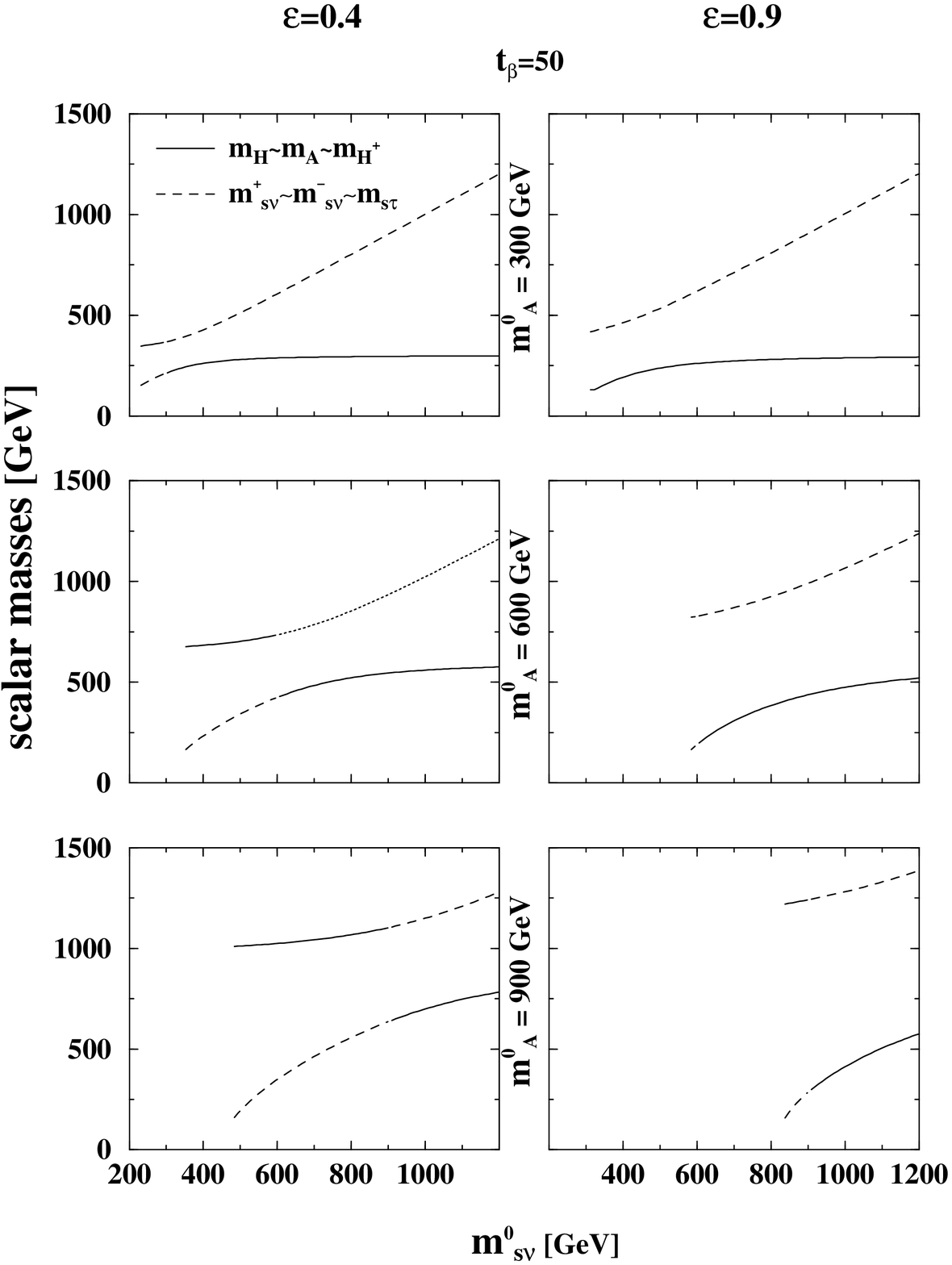,height=8cm,width=16cm,bbllx=0cm,bblly=2cm,bburx=20cm,bbury=25cm,angle=0}
 \end{center}
\caption{Physical masses of 
the heavy CP-even Higgs ($m_H$), 
CP-odd Higgs ($m_A$), charged Higgs ($m_{H^+}$),  
CP-even tau-sneutrino ($m_{s \nu}^+$), CP-odd 
tau-sneutrino ($m_{s \nu}^-$) and the stau ($m_{s \tau}$), as a function 
of the ``bare'' tau-sneutrino mass ($m_{s \nu}^0$), for $t_\beta=50$, 
for $m_A^0=300,~600$ or $900$ GeV and for $\varepsilon=0.4$ (left figures)
and $\varepsilon=0.9$ (right figures). $\epsilon$ is defined 
in (\ref{b3toeps}).}  
\label{figtypeB2}
\end{figure*}

\begin{enumerate}
  
\item A non-vanishing $b_3$ can generate a radiative 
(one-loop) tau-neutrino mass.   
The laboratory limit on the tau-neutrino mass allows, however,   
the quantity $\frac{b_3}{b_0} \sim \varepsilon$ to be 
of $\sim {\cal O}(1)$ \cite{davidson}.

\item The parameter $b_3$, or equivalently the 
quantity $\varepsilon \sim \frac{b_3}{b_0}$, can 
have important consequences 
on the 
CP-even and CP-odd Higgs-like scalar mass spectrum, see \cite{ourp1,ourp2}.
 In particular, 
$\varepsilon$ can drive the mass of the physical CP-even light Higgs 
below its 
present LEP2 lower bound which, for $m_A \gsim 200$ GeV,  
is roughly $m_h \gsim 110 $ GeV irrespective of $\tan\beta$ 
\cite{mhlim1}.$^{[12]}$\footnotetext[12]{This bound is 
applicable in the maximal 
stop mixing scenario with a typical SUSY squark masses of 
1 TeV. Note also that since $b_3 \neq 0$ 
the $hZZ$ coupling is smaller than its value in the RPC case 
leading to a smaller $e^+e^- \to Zh$ production rate. 
Thus, the limits reported in \cite{mhlim1} 
should be slightly relaxed in the type B RPV scenario.}
Also,
a non-zero $\varepsilon$ can 
give rise to negative eigenvalues (i.e., to the physical square masses) 
for the CP-even and CP-odd mass matrices $M_E^2$ and $M_O^2$ in 
(\ref{me2}) and (\ref{mo2}), depending on the values of the rest of 
the type B 
parameter space, i.e., on  $m_A^0$, $m_{s \nu}^0$ and $t_\beta$.

Therefore,  
in what follows, we will vary the parameters 
$\left\{m_A^0,m_{s \nu}^0,t_\beta,\varepsilon \right\}$
subject to the existing LEP2 lower bound on the light Higgs mass and 
to the requirement that 
$m_{\tilde\nu_{+}^\tau},~m_{\tilde\nu_{-}^\tau}$ and 
$m_{\tilde\tau^+}$ are $ > 150$ GeV. 

Since the light Higgs mass is very sensitive to higher order corrections 
to the $2 \times 2$ Higgs block in $M_E^2$, as in \cite{ourp1,ourp2}, 
in order to derive realistic exclusion regions for the parameter space  
$\left\{m_A^0,m_{s \nu}^0,t_\beta,\varepsilon \right\}$ through the 
requirement $m_h \gsim 110 $ GeV, we include 
the dominant higher order corrections (coming from 
the $t - \tilde t$ sector) to the $(\xi_d^0,\xi_u^0)$
block in $M_E^2$, following the approximate formulae given in 
\cite{mhcorrections} and taking the maximal mixing scenario with a 
typical squark mass of $m_{\tilde q} = 1$ TeV.  

\item A non-vanishing $\varepsilon$ can also alter the cross-section 
for $ZZ$ and $WW$ pair production through $s$-channel exchanges 
of the CP-even scalars $\Phi_E$ \cite{ourp1,ourp2}.
The measured $ZZ$ and $WW$ cross-sections in LEP2 can thus be used 
to place limits on $\varepsilon$ as a function of  
$\left\{m_A^0,m_{s \nu}^0,t_\beta \right\}$. These limits, however, 
can be evaded if the $\tilde\nu^\tau_{\pm}e^-e^+$ trilinear 
RPV coupling $\lambda_{131}$ is assumed 
small enough (see \cite{ourp1,ourp2}). We will, therefore, not consider
such limits below.

\end{enumerate}

\item{II.} Since $b_3$ is not a flavor changing parameter, the transition 
between down-quarks of different generations, i.e., between 
the external down-quarks $d_I \to d_J$, is necessarily 
driven by a $\lambda^\prime$ coupling with the appropriate 
non-diagonal indices (disregarding flavor changing transitions 
due to small non-diagonal CKM elements). 
Thus, the type B RPV one-loop effect in $Z \to d_I \bar d_J$ 
is necessarily proportional to either $b_3 \lambda^\prime_{3 I J}$ or
$b_3 \lambda^\prime_{3 J I}$.

In particular, for $Z\to b \bar s$ we find that the dominant 
contribution is attributed to the type B1 exchanges of the charged 
scalars and it arises when $\lambda^\prime_{3 32} \ne 0$. The only other 
possible index combination for $Z\to b \bar s$, which is 
$\lambda^\prime_{323} \ne 0$, yields a much smaller branching ratio.
This enhancement for the (332) index combination  
can be traced to the fact that, for this particular combination, 
the 
charged scalar amplitude involves also a top-quark exchange, thus gaining 
a factor of $m_t/m_c$ compared to 
the $\lambda^\prime_{3 23} \ne 0$ case (which 
involves a charm-quark exchange in the loops). 

\item{III.} In the limit $\varepsilon \to 0$ the type B2 effect vanishes.
However, since $\varepsilon \to 0$ causes the 
charged Higgs sector to decouple from the stau sector 
and since the RPC MSSM Higgs sector is similar to the 2HDM of type II, 
the type B1 contribution approaches that of the type II 2HDM 
in this limit. Thus, 
for $\varepsilon \to 0$, the type B1 RPV effect will be   
proportional 
to the off-diagonal CKM elements as in the case of
the type II 2HDM discussed in section III.
   
\item{IV.} In the numerical analysis below 
we will set $\lambda^\prime_{3 32} = 1$, 
while all other lambda's with different index combinations are set to zero. 
The experimental limit on this coupling, derived from 
$R_\ell =\Gamma(Z\to {\rm hadrons})/\Gamma(Z\to \ell \bar\ell)$ 
\cite{hepph9503264},
is (at the $2 \sigma$ level) 
$\lambda^\prime_{3 32}=0.45$ for squark masses of $\sim 100$ GeV  
, while $\lambda^\prime_{3 32} =1$ is allowed for 
squark masses $\gsim 650$ GeV. The perturbativity bound on this 
coupling is $\lambda^\prime_{3 32} =1.04$ \cite{hepph9906224}. 
Thus, we will assume that the squarks are heavy enough to allow   
$\lambda^\prime_{3 32}$ to lie near its perturbativity limit 
(recall that  
no squarks are involved in the type B RPV contribution 
to $Z \to b \bar s$). 
   
\end{description}

In Fig.~\ref{figtypeB1} we show
$BR(Z\to b \bar s + \bar b s)$ as a function of the ``bare'' tau-sneutrino 
mass $m_{s\nu}^0$ 
(i.e., what would be its mass in the RPC limit), for various 
possible values of $m_A^0$, $\varepsilon$ and for 
$t_\beta=3$ (left side) and $t_\beta=50$ (right side) \cite{curves}.
Evidently, $BR(Z\to b \bar s + \bar b s)$ is much larger in the 
high $\tan\beta$ scenario and it drops with $m_{s\nu}^0$.

The masses
of the heavy CP-even Higgs, CP-odd Higgs and charged Higgs 
as well as the CP-even, CP-odd 
tau-sneutrino and the stau particles are depicted in 
Fig.~\ref{figtypeB2}, for $t_\beta=50$ and for 
the same combinations of $\varepsilon$ and $m_A^0$ that are used 
in Fig.~\ref{figtypeB1}.
We note that 
in the limit $(m_A^0)^2 \gg m_Z^2$ (applicable 
to the values of $m_A^0$ in Figs.~\ref{figtypeB1} and \ref{figtypeB2})
one has $m_H \sim m_A \sim m_{H^+}$ and if in addition 
$(m_{s \nu}^0)^2 \gg m_Z^2$, then also the CP-even, CP-odd tau-sneutrinos 
and the stau are roughly degenerate.   
Thus, 
only two curves are shown in each plot in Figs.~\ref{figtypeB2}, which 
are sufficient 
to approximately describe all these six scalar masses.

Fig.~\ref{figtypeB2} shows that at some instances, 
the Higgs-like and slepton-like scalar masses exhibit a 
discontinuous jump, at 
which point they ``switch'' identities. 
This phenomena is caused by the 
particular 
dependence of the physical scalar masses on the ``bare'' masses 
$m_A^0$ and $m_{s \nu}^0$ in the presence of $\varepsilon \neq 0$. 
In particular, 
the corrections to the ``bare'' scalar masses due to a non-vanishing 
$b_3$ term are proportional 
to factors of $\left[ (m_A^0) - (m_{s \nu}^0) \right]^{-1}$ (for more 
details see \cite{ourp1,ourp2}), thereby
changing sign at the turning points.
Moreover, the off-diagonal elements of the rotation matrices 
$R_E,~R_O$ and $R_C$, which are responsible for the slepton-Higgs mixings, 
are also inversely proportional to factors of 
$\left[ (m_A^0) - (m_{s \nu}^0) \right]$, therefore, enhancing 
the type B RPV effect as $m_A^0$ approaches $m_{s \nu}^0$ as can be 
seen in Figs.~\ref{figtypeB1}.

\begin{table*}[htb]
\begin{center}
%\begin{tabular}{||c|c|c||}
\begin{tabular}{c|c|c}
%\hline
%%%%%%%%%%%%%%%%%%%%%%%%%%%%%%%%%%%%%%%%%%%%%%%%%%%%
{\bf model}  & {\bf scalars in the loops} & \boldmath{$BR(Z\to b \bar s)$} \\ 
~ & ~ & \\
\hline
~ & ~ & \\
SM & $W$-boson (no scalars) & $10^{-8}$ \\
~ & ~ & \\
%\hline
2HDMII & charged Higgs & $10^{-10}$ \\
~ & ~ & \\
%\hline
T2HDM & charged Higgs &  $10^{-8}$ \\
~ & ~ & \\
%\hline
SUSY with $\tilde t -\tilde c$ mixing & $\tilde t -\tilde c$ admixtures & 
$10^{-8}$ \\
~ & ~ & \\
%\hline
SUSY with $\tilde b -\tilde s$ mixing & $\tilde b -\tilde s$ admixtures  & 
$10^{-6}$  \\
~ & ~ & \\
%\hline
SUSY with trilinear $R_P$-violation & 
squarks and sleptons & $10^{-10}$  \\
~ & ~ & \\
%\hline
SUSY with bilinear $R_P$-violation & slepton-Higgs admixtures &
$10^{-6}$ \\
~ & ~ & \\
%\hline \hline
\end{tabular} 
\caption{The best case values for 
the branching ratio of $Z \to b \bar s$ 
for each of the six models considered in this paper upon imposing 
the available experimental limits on the relevant parameter space of each 
of them. The SM prediction is also given.}  
\label{sumtab}
\end{center}
\end{table*}

To summarize this section, with a large $\tan\beta$, 
a $BR(Z\to b \bar s + \bar b s) \sim {\cal O}(10^{-6})$
is possible within the type B scenario, e.g.,  for $40\%$ lepton 
number violation 
in the SUSY scalar potential ($\varepsilon=0.4$) and if the 
sleptons masses lie around $\sim 200$ GeV. 
For a heavier slepton spectrum a larger $\varepsilon$ 
is required in order to push the branching ratio to the $10^{-6}$ level. 

It should also be emphasized that since $BR(Z\to b \bar s + \bar b s)$ 
is dominated by the $\lambda^\prime_{332}$, 
the decay $Z \to b \bar s$ is an efficient probe 
of this specific flavor changing trilinear RPV coupling.

\section{Experimental feasibility}

In this section we will very briefly comment about the 
feasibility of observing 
(or achieving a limit) a signal of $Z \to b \bar s$ with a branching ratio 
of order $10^{-6}$, at a Linear Collider producing $10^9$ Z-bosons. 

Such a signal should appear in the detector as an event 
with one b-jet and one light-jet (assuming no distinction 
is made between ${\rm light}=d,u$ or $s$-quarks). 
In the spirit of the analysis made with the 1993 and 1994 LEP data 
on $Z \to b \bar s$ \cite{delphi}, one 
defines $\epsilon_q^B$ and $\epsilon_q^L$ to be the
efficiencies that a quark (or anti-quark) of 
flavor $q$ is tagged as a $b$-jet ($B$) and light-jet ($L$), respectively.
Thus, the key efficiency parameters for the detection of $Z \to b \bar s$ are 
$\epsilon_b^B,~\epsilon_{\rm light}^L$ and $\epsilon_b^L$, where the latter 
represent the probability that a $b$-jet is identified as a light-jet and 
is important for controlling the dominant background (to the 
$Z \to B+L$ 
signal) coming from $Z \to b \bar b$. Note, that due to the expected smallness 
of the "Purity" parameters $\epsilon_{\rm light}^B,\epsilon_c^B$ and 
$\epsilon_c^L$ (see \cite{delphi}), 
the background to $Z \to B+L$  
caused by the SM $Z\to d \bar d,~s \bar s,~u \bar u,~c \bar c$ decays 
will be sub-dominant.     

With $10^9$ Z-bosons, the expected 
number of events coming from $Z \to b \bar s$ (i.e., from new physics) 
and identified as $Z \to B+L$, is:

\begin{eqnarray}
S \sim 10^9 \times \epsilon_b^B \epsilon_{\rm light}^L 
BR(Z \to b \bar s) \label{ns}~.
\end{eqnarray}

\noindent Similarly, the expected number of background 
 $Z \to B+L$ events coming from the SM decay
$Z \to b \bar b$ is:

\begin{eqnarray} 
B \sim 10^9 \times \epsilon_b^B \epsilon_b^L BR(Z \to b \bar b) ~\label{nb}.
\end{eqnarray}

\noindent Using (\ref{ns}) and (\ref{nb}), the expected
statistical significance, $S/\sqrt{B}$,
of the new physics signal $Z \to b \bar s$, with 
a branching ratio of order $10^{-6}$, can reach beyond
the 3-sigma level for
$\epsilon_b^B \sim 0.6 -0.8$, $\epsilon_{\rm light}^L \sim 0.3-0.5$ and
$\epsilon_b^L \sim {\cal O}(10^{-4})$. These values require an improvement
to the 1993 and 1994 analysis \cite{delphi}, by a factor of 2-3 for
$\epsilon_b^B$ and $\epsilon_{\rm light}^L$ and by an order of magnitude
for $\epsilon_b^L$.
With the expected advancement in the jet-tagging methods, in particular,
for two-body decays of the Z-boson, these required values for the efficiency 
parameters above should be well within the reach of the future Linear 
Collider. 

We can also get a clue about how low one can go in the value 
(or limit) of $BR(Z\to b \bar s)$ with $10^9$ Z-bosons, from the fact that 
the LEP preliminary results \cite{delphi} achieved 
$BR(Z\to b \bar s) < {\cal O}(10^{-3})$ with ${\cal O}(10^{6})$ Z-bosons.
Scaling this limit, especially with the expected advance in b-tagging 
and identification of non-b jets methods, an ${\cal O}(10^{-6})$ branching 
ratios should be easily attained at a Giga-Z factory.

\section{Summary}

We have re-examined the flavor changing radiative decays 
of a $Z$-boson to a pair of down-quarks, $Z \to d_I \bar d_J$, with
$I \neq J$. These Z-decay channels may prove useful in 
searching for new flavor physics beyond the SM at the TESLA collider, or any 
other future collider, which 
may be designed to run on the Z-pole with high luminosities, 
thus accumulating more than $10^9$ on-shell Z-bosons. 
With advances in technology, e.g., improved b-tagging efficiencies,
the flavor changing decay $Z \to b \bar s$ - most likely the easiest 
to detect among the flavor changing hadronic Z-decays - may be accessible 
to a Giga-Z option even for branching ratios 
as small as $BR(Z \to b \bar s) \sim 10^{-7} -10^{-6}$.   

The $d_I \to d_J$ transition was assumed to be 
generated at one-loop through flavor violation in interactions 
between scalars and fermions. 

A complete analytical derivation of the width $\Gamma(Z \to d_I \bar d_J)$ 
is presented using the form factor approach for the 
$Zd_I\bar d_J$ interaction vertex.
These form factors are evaluated for the complete set of 
scalar-fermion one-loop exchanges with generic 
scalar-fermion flavor-violating couplings.

This prescription is then applied to the decay $Z \to b \bar s$ 
in six beyond the SM model scenarios  
for flavor-violation in the scalar sector:

\begin{enumerate}
\item Two Higgs doublet models with non-standard charged-Higgs couplings 
to quarks:
\begin{itemize}
\item A two Higgs doublet model of type II (2HDMII). 
\item A two Higgs doublet model "for the top-quark" (T2HDM).
\end{itemize}
\item Supersymmetry with flavor-violation in the squark sector:
\begin{itemize}
\item Supersymmetry with stop-scharm mixing.
\item Supersymmetry with sbottom-sstrange mixing. 
\end{itemize}
\item Supersymmetry with flavor-violation from R-parity violating 
interactions:
\begin{itemize}
\item Supersymmetry with trilinear R-parity violation.
\item Supersymmetry with trilinear and bilinear R-parity violation. 
\end{itemize}
\end{enumerate}

Folding in the existing experimental limits on 
the relevant parameter space of each of these models, we calculated the 
branching ratio for the decay $Z \to b \bar s$. The highlights of
our results are summarized in Table \ref{sumtab}.
In particular, we find that two Higgs doublet models with flavor 
violation originating from charged 
scalar interactions with fermions
are expected to yield an extremely small $BR(Z \to b \bar s)$; 
smaller than the SM prediction and smaller than the reach of 
a Giga-Z $\ell^+ \ell^-$ collider.
Thus, a signal of $Z \to b \bar s$ in TESLA will be inconsistent 
with the underlying mechanisms for flavor violation in these two Higgs 
doublets model and will, therefore, rule out these options.   

The same conclusions can be drawn in the stop-scharm mixing
and the trilinear R-parity violation SUSY scenarios.
On the other hand, SUSY with 
mixings between the bottom and strange-type squarks 
and/or mixings between sleptons and Higgs fields (bilinear R-parity violation)
both of which may originate from the soft SUSY breaking sector,
can drive the $BR(Z \to b \bar s)$ to the $10^{-6}$ level for
large $\tan\beta$ values. 
This enhancement is typical to these two flavor-violating SUSY scenarios 
if there are large mass-splittings between the scalars exchanged in the
loops due
to a GIM-like cancellation which is operational in 
the scalar mass-matrices and is, therefore, less effective as 
the scalar masses depart from degeneracy.

A $Z \to b \bar s$ signal in a Giga-Z TESLA or any other collider 
may, therefore, be a good indication for the underlying dynamics of these two
flavor-violating SUSY scenarios and, if interpreted in that way, 
will provide for evidence of an hierarchical 
structure in the mass spectrum of the SUSY scalar sector.

\begin{acknowledgments}
G.E. and A.S. thank the
U.S.-Israel Binational Science Foundation. G.E. also thanks 
the Israel Science
Foundation and the Fund for Promotion of Research at the
Technion for partial support.
This work was also supported in part by US DOE Contract Nos.
DE-FG02-94ER40817 (ISU) and DE-AC02-98CH10886 (BNL).
\end{acknowledgments}

\appendix

\section{One-loop form factors}
%\section*{Appendix - one-loop form factors}

In this appendix we give the two-point and three-point one-loop form factors 
which are defined by the one-loop momentum integrals as follows \cite{1loop1}:

\begin{widetext}
\begin{eqnarray}
C_{0};~C_{\mu};~C_{\mu \nu} 
\left(m_1^2,m_2^2,m_3^2,p_1^2,p_2^2,p_3^2 \right) & \equiv& 
\int \frac{d^4 q}{i \pi^2}
\frac{1;~q_\mu;~q_\mu q_\nu}{ 
\left[q^2 -m_{1}^2 \right] \left[ (q+p_1)^2 -m_{2}^2 \right]
\left[ (q-p_3)^2 -m_{3}^2 \right] } ~,\\
\tilde C_{0};~\tilde C_{\mu} 
\left(m_1^2,m_2^2,m_3^2,p_1^2,p_2^2,p_3^2 \right) & \equiv& 
\int \frac{d^4 q}{i \pi^2}
\frac{q^2;~q^2 q_\mu}{ 
\left[q^2 -m_{1}^2 \right] \left[ (q+p_1)^2 -m_{2}^2 \right]
\left[ (q-p_3)^2 -m_{3}^2 \right] } ~,\\
B_{0};~B_{\mu} \left(m_1^2,m_2^2,p^2 \right) &\equiv& 
\int \frac{d^4 q}{i \pi^2}
\frac{1;~q_\mu}{ 
\left[q^2 -m_{1}^2 \right] \left[ (q+p)^2 -m_{2}^2 \right] } ~,
\end{eqnarray}
\end{widetext}

\noindent where $\sum_i p_i=0$ is to be understood above.

The coefficients $B_x$ with $x \in 0,1$, $C_x$ with 
$x \in 0,11,12,21,22,23,24$ and $\tilde C_x$ with $x \in 0,11,12$ 
are then defined through the following relations \cite{1loop2}:

\begin{eqnarray}
B_\mu &=& p_{\mu} B_1 ~,\\
C_\mu &=& p_{1\mu} C_{11} + p_{2\mu} C_{12} ~,\\
\tilde C_\mu &=& p_{1\mu} \tilde C_{11} + p_{2\mu} \tilde C_{12} ~,
\end{eqnarray}

\n and
\begin{equation}
C_{\mu \nu} = p_{1 \mu} p_{1 \nu} C_{21} +
p_{2 \mu} p_{2 \nu} C_{22} + \left\{ p_{1}p_{2}\right\}_{\mu \nu} C_{23} 
+ g_{\mu \nu} C_{24} ~,
\end{equation}

\noindent where $\left\{ab \right\}_{\mu \nu} 
\equiv a_\mu b_\nu + a_\nu b_\mu$.


\begin{thebibliography}{99}

\bibitem{isidori} For a recent review on rare $K,~D$ and
$B$ decays, see e.g.: G. Isidori, hep-ph/0110225. 

\bibitem{mele} For recent reviews on the 
experimenal and theoretical status of rare
top decays, see e.g. : B. Mele, hep-ph/0003064 and G.G. Hanson,
hep-ex/0111058, respectively. 

\bibitem{SM} 
M. Clements {\it et al.}, Phys. Rev. {\bf D27}, 570 (1983);
V. Ganapathi {\it et al.}, Phys. Rev. {\bf D27}, 579 {1983};
W.-S. Hou, N.G. Deshpande, G. Eilam and A. Soni, Phys. Rev. Lett.
{\bf 57}, 1406 (1986);
 J. Bernabeu, M.B. Gavela and A. Santamaria,
Phys. Rev. Lett. {\bf 57}, 1514 (1986).

\bibitem{2HDMIIref}
C. Busch, Nucl. Phys. {\bf 319}, 15 (1989);
W.-S. Hou and R.G. Stuart, Phys. Lett. {\bf B226}, 122 {1989};
B. Grzadkowski, J.F. Gunion and P. Krawczyk, 
Phys. Lett. {\bf 268}, 106 (1991). 


\bibitem{bsmixold} B. Mukhopadhyaya and A. Raychaudhuri, Phys. Rev.
{\bf D39}, 280 (1989); see also M. J. Duncan,
Phys. Rev. {\bf D31}, 1139 (1985);
F. Gabbiani, J.H. Kim and A. Masiero,
Phys.Lett. {\bf B214}, 398 (1988).

\bibitem{shemtob} M. Chemtob and G. Moreau,
Phys. Rev. {\bf D59}, 116012 (1999).

\bibitem{others} 
W. Buchm\"{u}ller and M. Gronau, Phys. Lett. {\bf B220}, 641 (1989);
G.T. Park and T.K. Kuo, 
Phys. Rev. {\bf D42} 3879, (1990); 
M.A. Perez and M.A. Soriano, Phys. Rev. {\bf D46}, 284 (1992); 
J. Roldan, F.J. Botella and J. Vidal, Phys. Lett. {\bf B283}, 389 (1992);
X.-L. Wang, G.-R. Lu and Z.-J. Xiao, Phys. Rev. {\bf D51}, 4992 (1995).


\bibitem{delphi} J. Fuster, F. Martinez-Vidal and P. Tortosa,
preprint DELPHI 99-81 CONF 268, June 1999. Work presented at: 
``International Europhysics Conference on High-Energy Physics, Tampere, 
Finland, 15-21, July 1999-IOP, Bristol, 1999. Paper can be downloaded 
from: http://documents.cern.ch/cgi-bin/setlink?base=preprint\&categ=cern\&id=open-99-393.  

\bibitem{pdg} D.E. Groom {\it et al.} (Particle Data Group),
Euro. Phys. J. {\bf C15}, 1 (2000).

\bibitem{ZinLHC} A.D. Martin, R.G. Roberts, W.J. Stirling and R.S. Thorne,
Eur. Phys. J {\bf C14}, 133 (2000). 
 
\bibitem{ZinTESLA} J.A. Aguilar-Saavedra {\it et al.},
(ECFA/DESY LC Physics Working Group), hep-ph/0106315.

\bibitem{mucol1} C.M. Ankenbrandt, hep-ph/9901022.

\bibitem{mucol2} D. Atwood, L. Reina and A. Soni,
Phys.\ Rev.\ Lett.\ {\bf 75}, 3800 (1995);
M. Sher, Phys.\ Lett.\ {\bf B487}, 151 (2000). 

\bibitem{yukawaref} See e.g., D. Atwood, S. Bar-Shalom, G. Eilam and 
A. Soni, Phys.\ Rep.\ {\bf 347}, 1 (2001) and references therein. 

\bibitem{froggat} C.D. Froggatt, R.G. Moorhouse, I.G. Knowles, 
Nucl.\ Phys.\ {\bf B386}, 63 (1992). 

\bibitem{das} A.K. Das and C. Kao, Phys.\ Lett.\ {bf B372}, 106 (1996). 

\bibitem{9810552} K. Kiers, A. Soni and G.-H. Wu,
Phys.\ Rev.\ {\bf D59}, 096001 (1999).


\bibitem{2HDMIIlim} M. Ciuchini, G. Degrassi, P. Gambino 
and G.F. Giudice, Nucl.\ Phys.\ {\bf B527}, 21 (1998). See also,
P. Krawczyk and S. Pokorski, Phys.\ Rev.\ Lett.\ {\bf 60}, 182 (1988);
J.L. Hewett, Proceedings of the SLAC Summer Inst. on Particle 
Physics: Spin Structure in High Energy Processes, Stanford, CA, 1993, 
hep-ph/9406302;
Y. Grossman, H. Haber and Y. Nir, Phys.\ Lett.\ {\bf B357}, 630 (1995);
J.L. Hewett and J.D. Wells, Phys.\ Rev.\ {\bf D55}, 5549 (1997). 
 

\bibitem{t2hdmlim} G.-H. Wu and A. Soni, 
Phys.\ Rev.\ {\bf D62}, 056005 (2000);
K. Kiers, A. Soni and G.-H. Wu, 
Phys.\ Rev.\ {\bf D62}, 116004 (2000). 

\bibitem{misiak1} M. Misiak, S. Pokorski and J. Rosiek, 
Adv.\ Ser.\ Direct.\ High Energy Phys.\ {\bf 15}, 795 (1998). 
%hep-ph/9703442; 

\bibitem{yuan} J.L. Diaz-Cruz, H.-J. He and C.-P. Yuan, hep-ph/0103178.


\bibitem{rosiek} J. Rosiek, 
Phys.\ Rev.\ {\bf D41}, 3464 (1990), unpublished Erratum in hep-ph/9511250.

\bibitem{deltalimits}  
F. Gabbiani, E. Gabrielli, A. Masiero and L. Silvestrini,
Nucl.\ Phys.\ {\bf B477}, 321 (1996);
F. Borzumati, C. Greub and T. Hurth, Phys.\ Rev.\ {\bf D62}, 075005 (2000);
T. Besmer, C. Greub and T. Hurth, Nucl.\ Phys.\ {\bf B609}, 359 (2001).

\bibitem{curves} The 
curves in the figures end abruptly when the values of the parameter 
space involved 
is not compatible with our imposed restrictions from experimental data 
on the relevant physical quantities.  

\bibitem{rpvreview} For reviews on R-parity violation see e.g., 
D.P. Roy, Pramama, J.\ Phys.\ {\bf 41}, S333 (1993), also in 
hep-ph/9303324;
G. Bhattacharyya, Nucl.\ Phys.\ B (Proc. Suppl.) {\bf 52A}, 83 (1997);
H. Dreiner, in {\it Perspectives in Supersymmetry}, edited by G.L. Kane 
(World Scientific, Singapore, 1998), hep-ph/9707435;
P. Roy, "Seoul 1997, Pacific Particle Physics Phenomenology", 
talk given at APCTP Workshop: Pacific Physics Phenomenology (P4 97), 
Seoul, Korea, 1997, hep-ph/9712520;
S. Raychaudhuri, hep-ph/9905576. 

\bibitem{GH} Y. Grossman and H.E. Haber,
Phys.\ Rev.\ {\bf D59}, 093008 (1999).

\bibitem{morebterms} S. Roy and B. Mukhopadhyaya,
Phys.\ Rev.\ {\bf D55}, 7020 (1997);
M.A. Diaz, J.C. Romao and J.W.F. Valle,
Nucl.\ Phys.\ {\bf B524}, 23 (1998);
C.-h. Chang and T.-f. Feng, hep-ph/9908295;
B. Mukhopadhyaya and S. Roy,
Phys.\ Rev.\ {\bf D60}, 115012 (1999).

\bibitem{davidson} S. Davidson, M. Losada and N. Rius, Nucl.\ Phys.\
{\bf B587}, 118 (2000).

\bibitem{ref12} See e.g., K. Choi, E.J. Chun and K. Hwang,
Phys.\ Lett.\ {\bf B488}, 145 (2000).

\bibitem{ref8} M. Bisset, O.C.W. Kong, C. Macesanu and 
L.H. Orr, Phys.\ Rev.\ {\bf D62}, 035001 (2000).

\bibitem{neutrinomass} See e.g., H.-P. Nilles and N. Polonsky,
Nucl.\ Phys.\ {\bf B484}, 33 (1997);
B. Mukhopadhyaya, S. Roy and F. Vissani,  Phys.\ Lett.\ {\bf
B443}, 191 (1998);
M. Hirsch, M.A. Diaz, W. Porod, J.C. Romao and J.W.F. Valle, 
Phys.\ Rev.\ {\bf D62}, 113008 (2000).

\bibitem{basis} J. Ferrandis, Phys.\ Rev.\ {\bf D60}, 095012
(1999).

\bibitem{limlamlam} For a list of updated bounds on the 
$\lambda^\prime \lambda^\prime$ products that enter the radiative 
decay $Z \to b \bar s$ see G. Bhattacharyya, D. Chang, C.-H. Chou and 
W.-Y. Keung, Phys.\ Lett.\ {\bf B493}, 113 (2000). 

\bibitem{bsglamlamlim} B. de Carlos and P.L. White, 
Phys.\ Rev.\ {\bf D55}, 4222 (1997);
T. Besmer and A. Steffen, Phys.\ Rev.\ {\bf D63}, 055007 (2001).  


\bibitem{davidsonnew} Y. Grossman and H.E. Haber, 
Phys.\ Rev.\ {\bf D63}, 075011 (2001);
S. Davidson and M. Losada, hep-ph/0010325.


\bibitem{ourp1} S. Bar-Shalom, G. Eilam and B. Mele, 
Phys.\ Lett.\ {\bf B500}, 297 (2001).

\bibitem{ourp2} S. Bar-Shalom, G. Eilam and B. Mele, 
Phys.\ Rev.\ {\bf D64}, 095008 (2001).

\bibitem{mhlim1} See e.g., talks given by Tom Junk at
the ``LEP Fest'', 10 October 2000, and by P. Teixeira-Dias, 10 July 2001, in
http://lephiggs.web.cern.ch/LEPHIGGS/talks.

\bibitem{mhcorrections} S. Heinemeyer, W. Hollik 
and G. Weiglein, hep-ph/0002213.

\bibitem{hepph9503264} G. Bhattacharyya, J. Ellis and K. Sridhar,
Mod.\ Phys.\ Lett.\ {\bf A10} , 1583 (1995).

\bibitem{hepph9906224} B. Allsnsch {\it et al.}, hep-ph/9906224, 
edited by H. Dreiner.  

\bibitem{1loop1} For the FF package that was used for numerical evaluation 
of the loop integrals see: G.J. van Oldenborgh, 
Comput.\ Phys.\ Commun.\ {\bf 66}, 1 (1991).
For the algorithms used in the FF package see: G.J. van Oldenborgh
and J.A.M. Vermaseren, Z.\ Phys.\ {\bf C46}, 425 (1990).    

\bibitem{1loop2} G. Passarino and M. Veltman, 
Nucl.\ Phys.\ {\bf B160}, 151 (1979).

\end{thebibliography}
\end{document}